\documentclass[prb,twocolumn,preprintnumbers,amsmath,amssymb]{revtex4-1}
\pdfoutput=1
\usepackage{graphicx}
\usepackage{color}
\usepackage{bm}

\begin{document}
\author{M. Wimmer}
\author{A. R. Akhmerov}
\affiliation{Instituut-Lorentz, Universiteit Leiden, P.O. Box 9506, 2300 RA Leiden, The Netherlands}

\author{F. Guinea}
\affiliation{Instituto de Ciencia de Materiales de Madrid, CSIC, Sor
Juana In\'es de la Cruz 3, E28049 Madrid, Spain}

\title{Robustness of edge states in graphene quantum dots}

\date{March, 2010}

\begin{abstract}
  We analyze the single particle states at the edges of disordered
  graphene quantum dots. We show that generic graphene quantum dots support a
  number of edge states proportional to circumference of the dot over
  the lattice constant. Our analytical theory agrees well with numerical
  simulations. Perturbations breaking
  electron-hole symmetry like next-nearest neighbor hopping or edge
  impurities shift the edge states away from zero energy but do not
  change their total amount. We discuss the possibility of detecting
  the edge states in an antidot array and provide an upper bound on
  the magnetic moment of a graphene dot.
\end{abstract}

\maketitle

\section{Introduction}

The experimental discovery\cite{Netal04,Netal05} of graphene, a
monolayer of carbon atoms, has opened room for new electronic
devices (for reviews, see Refs.~\onlinecite{GN07,ACP07,NGPNG09}). A
peculiarity of finite graphene sheets is the existence of electronic
states localized at the boundary, so-called \emph{edge states}.

A crystallographically clean zigzag edge was theoretically
predicted to sustain zero-energy edge states
\cite{K94,FWNK96,NFDD96}. Later, it was shown\cite{AB08} that any
generic graphene boundary not breaking electron-hole subband
(sublattice) symmetry also supports these zero energy edge states.
Similar states exist at zigzag edges of graphene
bilayers\cite{CPLNG08,SMMB08}, and in other multilayered graphene
systems \cite{EVCastro2008}. Experimentally, these states were
observed in STM experiments near monatomic steps on a graphite
surface\cite{Netal05b,KFEKK05,Netal06}.

The presence of large number of localized states is important for
the predicted edge magnetism in graphene
nanoribbons\cite{FWNK96,NIF98}, a topic that has recently seen
renewed interest in the context of graphene
spintronics\cite{SCL06,FP07,Ezawa07,YK08,Wimmer2008}. Apart from edge
magnetism, interacting edge states may also result in other
correlated ground states\cite{WSG08,WSSG08,RYL09}.

Edge states also play a role in confined geometries,  when the edge
to area ratio is large enough so that the electronic properties of
the boundary may become dominant. One example for  such a geometry
are graphene quantum dots that have been under intense experimental
study recently\cite{Petal08,Setal08,Stampfer2008,Guttinger2009,Setal09b,Getal10},
with quantum dot sizes in the range from a few tens of nanometers to
micrometers. Another example are antidot arrays that have been
subject of several theoretical studies\cite{Shima93,Pedersen08,Vanevic09,Fuerst09} and have also been
realized experimentally\cite{Shen2008,Eroms2009,Bai10,Balog10}.

Different edges have been observed in
graphite\cite{Cancado2004,KFEKK05,Netal05b,Netal06} and
graphene\cite{Vetal08,Jetal09,Getal09,Liu2009}. In particular, the
existence of boundaries with a long-range crystalline order in
exfoliated graphene has been questioned\cite{Casiraghi2009}. In
addition, the existence of unsaturated dangling bonds at edges makes
them reactive, and it is unclear how they are
passivated\cite{XL04,JSD06,CCPF08,HLWGD08}. Hence, it is likely that
graphene edges are perturbed and that the presence of edge distortions
has to be taken into account.

\begin{figure}
\includegraphics[width=\linewidth]{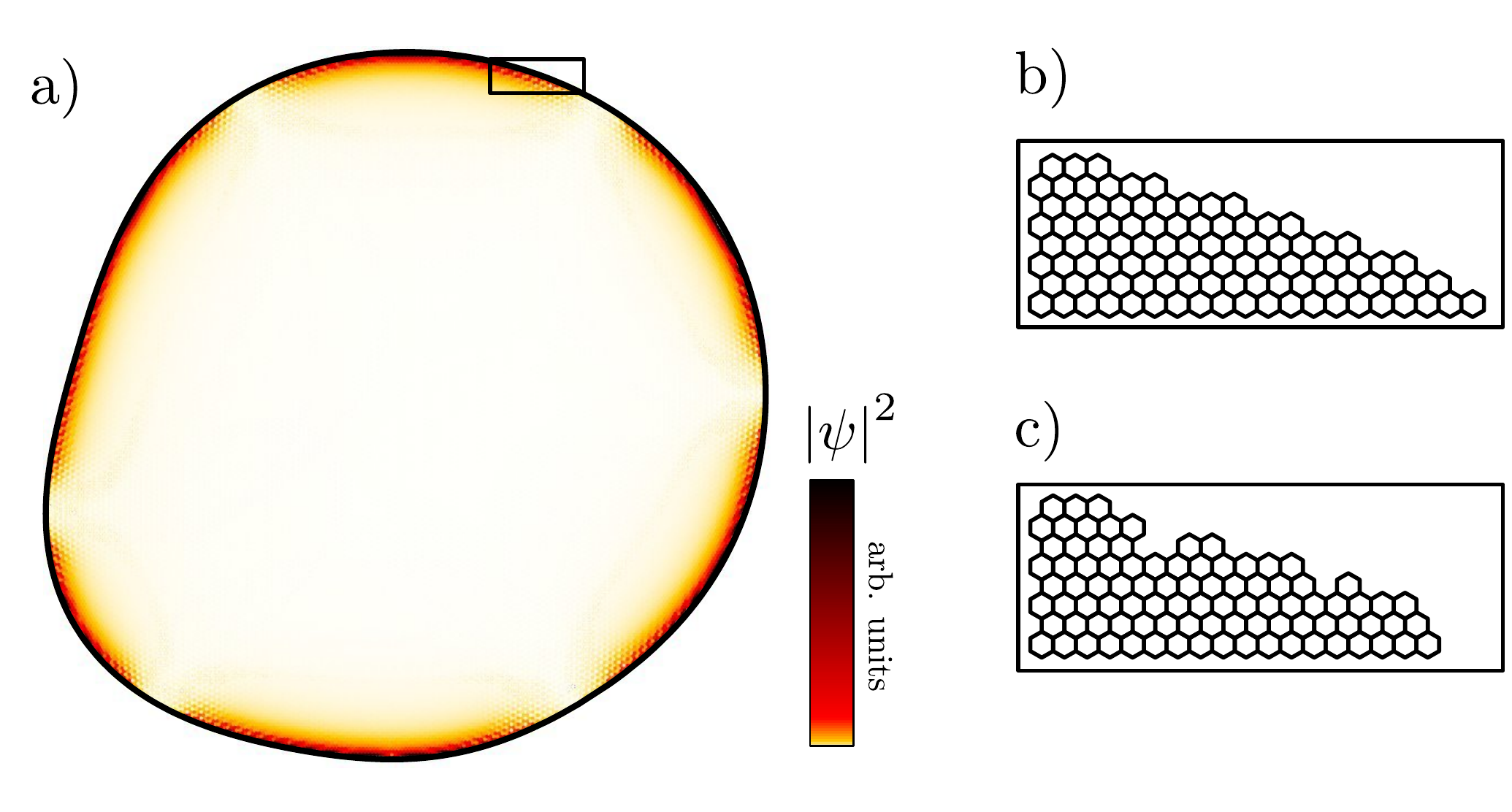}
\caption{A graphene quantum dot. The excess density of states due to
edge states is shown in a color plot\cite{footnote_colorplots},
as calculated for a quantum dot with a smooth boundary and no
particle-hole symmetry breaking perturbations (a). In general, edge
states are present both near a smooth boundary (b) and a boundary
with short range disorder (c).} \label{sketch}
\end{figure}

The aim of our paper is to show that edge states can  be expected in
realistic disordered quantum dots. We also analyse the particular
properties of edge states such as their number and compressibility.
We start the analysis in Section \ref{sec_analytic} by using the
theory of Ref.~\onlinecite{AB08} for a relation between the number of
edge states per unit length of a smooth boundary (see
Fig.~\ref{sketch}b) and the angle the boundary makes with respect to the
crystallographic axis. We extend the earlier results by calculating
the correction to the edge states number coming from the edge 
roughness (Fig.~\ref{sketch}c). Having the total number of edge
states and their momentum distribution we apply perturbation theory
to see how confinement energy and particle-hole symmetry breaking
terms in the Hamiltonian shift the edge states from zero energy.
Confinement energy spreads a delta function-like peak in the density
of states into a hyperbolic one. In contrast, particle-hole symmetry
breaking terms spread the edge states nearly homogeneously over a
band of finite width. For realistic dot sizes around tens of
nanometers we find the latter to be more important.

In Section \ref{sec_numeric} we perform numerical  simulations on
quantum dots of experimentally relevant sizes. These numerical
calculations confirm our analytic results. We also
study the magnetic field dependence of edge states in quantum dots.
Whereas magnetic field spectroscopy of energy levels has up to now
mainly been a useful tool to probe bulk states in graphene quantum
dots\cite{SESI08,Setal09b,Guttinger2009}, we show how to employ this
technique also to identify edge states. In addition, we study
the level statistics of edge states.

Finally in Section \ref{sec_discussion}, we calculate an upper bound
on the magnetic moment of a graphene dot due to edge state
polarization. We also give an upper bound on the relative weight of the
edge states with respect to the bulk states. By knowing the magnitude
of additional compressibility due to the edge states we estimate
parameters of an antidot lattice in which edge states would be visible
in SET experiments.

We conclude in Section \ref{Sec_conclusion}.

\section{Analytical calculation of the edge states density}\label{sec_analytic}

\subsection{Number of edge states}

The density of edge states per unit length was calculated for a
smooth edge in Ref.~\onlinecite{AB08}
\begin{equation}
 \frac{dN}{d l} = \frac{2}{3a}|\sin\phi|, \label{dndl}
\end{equation}
with $-\pi/6<\phi<\pi/6$ the angle boundary makes with a  nearest
armchair direction, and $a$ the lattice constant. This expression
is valid on the scales larger than the boundary roughness scale and another
scale $\delta(\phi)$ dependent on boundary structure. For most boundary orientations,
except the ones very close to armchair direction $\delta(\phi)\sim a$.
 Approximating the dot shape by a circle, and integrating
Eq.~\ref{dndl} along the whole perimeter of the dot, we get
\begin{equation}\label{numedgestates1}
 N=\int\limits_0^L\frac{dN}{dl}dL=\frac{4-2\sqrt{3}}{\pi}\times\frac{L}{a},
\end{equation}
with $L$ circumference of the dot and $a$ the lattice constant. This
density of states is the difference between total density of waves
evanescent away from the boundary and the number of conditions the wave
function must satisfy on the selected sublattice (see
Ref.~\onlinecite{AB08} for a more detailed description). If a small
fraction $\alpha$ of random outermost atoms of the smooth edge
oriented at angle $\phi$ with armchair direction is etched, the
number of conditions for the wave function on the minority sublattice
increases by
\begin{equation}
 \delta N = 2 \alpha \sin\phi.
 \label{numedgestates2}
\end{equation}
This leads to the reduction of the number of the edge states near an
edge with atomic scale disorder:
\begin{equation}\label{numedgestates3}
N'=N(1-2\alpha).
\end{equation}

Note that Eq.~\eqref{numedgestates3} only gives the local density of low energy edge states.
It should not be confused with Lieb's theorem,\cite{Lieb89} which connects the number of states with exactly zero energy
with the difference in the number of sublattice sites in a bipartitte sublattice. Lieb's theorem was applied to graphene in Refs.~\onlinecite{Shima93, Vanevic09, Ezawa10}, and for a disordered quantum dot geometry it predicts\cite{Vanevic09} number of zero-energy modes $\sim \sqrt{L}$. Our analysis shows that there will be $\sim L$ low energy edge states, although most of them do not lie at exactly zero energy. Hence, there is no contradiction
with Lieb's theorem.

\subsection{Edge state dispersion}\label{sec:dispersion}

There are two different mechanisms which give finite energy to
otherwise zero energy edge states: the overlap between edge states
on different sublattices, and terms breaking sublattice symmetry at
the edge. The dispersion resulting from these perturbations can be
calculated by applying degenerate perturbation theory, acting on the
wave functions $\psi_n$, belonging exclusively to $A$ or $B$ sublattice.
The long wavelength part of these wave functions is defined by the conformal
invariance of Dirac equation, so they can be approximated as plane waves
belonging to one of the six facets of the dot with well-defined boundary
condition, extended along the facet and decaying into the bulk.
These wave functions have longitudinal momenta
\begin{equation}
k_n\sim \frac{n}{R}\label{eq:momenta}
\end{equation}
approximately equally spaced due to phase space arguments.

We first estimate the energy dispersion due to edge state overlap,
or in other words by finite size effects. Particle-hole symmetry
prevents coupling between states on the same sublattice, so the
dispersion of edge states in a finite system can be calculated from
the matrix element between the edge states on different sublattices.
These states are separated from each other by a distance of an order
of the dot radius $R$ and their decay length away from the boundary
is proportional to difference $k$ between their momentum and the
momentum of the nearest Dirac point (Dirac momentum), so the energy
is
\begin{equation}
 E(k)\sim \frac{v_\text{F}}{R}e^{-kR},\label{eq:matrixelement}
\end{equation}
where $v_\text{F}$ is the Fermi velocity and we set $\hbar=1$.
We note that Eq.~\ref{eq:matrixelement} is very similar to the energy of edge
states in zigzag nanoribbons\cite{ABRB08}.
Substituting the value of momentum of the edge states from
Eq.~\eqref{eq:momenta}  into Eq.~\eqref{eq:matrixelement} we
calculate the density of edge states per unit energy
\begin{equation}\label{oneoverE}
\rho(E)\equiv \left|\frac{dn}{dE}\right|\sim\frac{1}{E}
\end{equation}

The atoms passivating the edge perturb the $\pi$-orbitals of carbon
atoms to which they are bound. This interaction breaks the effective
electron-hole symmetry of graphene around the Dirac point.
Next-nearest neighbor hopping is breaking this symmetry at the edges
as well\cite{PGN06,SMS06}, and it was shown to be equivalent to the
edge potential.\cite{Sasaki2009} The dispersion of the edge states
near a zigzag edge due to these two perturbations is
\begin{equation}
 E=(\Delta\epsilon-t')[\cos(K)-1/2],\;\; 2\pi/3<K<4\pi/3,\label{eq:ehbreakingdensity}
\end{equation}
where $K$ is the full momentum of the edge state,  $\Delta\epsilon$
is the average strength of the edge potential and $t'$ the
next-nearest neighbor hopping strength. Despite it is not
straightforward to generalize this equation to an arbitrary
orientation of the edge, the general effect of the electron-hole
symmetry breaking terms is to smear the zero energy peak in the
density of states into a band between energies of approximately 0 and
$E_0\equiv \Delta\epsilon - t'$ for the most localized states, while the
more extended states are near the Dirac energy.
The one dimensional van Hove singularity in the density of states
at $E=E_0$ will be smeared out, due to the presence
of a minimum decay length of the edge states when the orientation of
the boundary is not exactly zigzag.\cite{AB08}

The energy due to finite size of the dot given by
Eq.~\eqref{eq:matrixelement} is at best of an order of $E\sim
v_\text{F}/R\approx t a/R$. It is less than tens of millivolts for dots
above 10nm size. On the other hand the energy due to the edge
potentials and next-nearest neighbor hopping
(Eq.~\eqref{eq:ehbreakingdensity}) is likely to be around hundreds
of millivolts. Accordingly in realistic dots with edge potentials
and next-nearest neighbor hopping term edge states occupy the band
between the Dirac point and $E_0$ with approximately constant density
\begin{equation}
 \rho_\textrm{edge}=c(1-2\alpha)\frac{R}{a |E_0|}, \label{eq:edge_density}
\end{equation}
with $c=8-4\sqrt{3}\approx 1$.

\section{Numerical results} \label{sec_numeric}

In order to confirm the analytical results of the previous sections we
have performed numerical simulations of the energy spectrum of
graphene quantum dots with sizes relevant for experiments. In the
following we present results for a quantum dot with the shape of a
deformed circle\cite{dotshape} (c.f.~Fig.\ref{sketch}), characterized
by an average radius $R$. Although we focus on a particular quantum
dot here, we have found through numerical studies that the
characteristic features of our results are independent from the
details of the dot shape.

The numerical simulations are based on a tight-binding model of
graphene with Hamiltonian
\begin{equation}
H=-\sum_{i,j} t_{ij} c^\dag_i c_j + h. c.
\end{equation}
where the hopping $t_{ij}=t$ for nearest neighbors and $t_{ij}=t'$
for next-nearest neighbors\cite{NGPNG09}. The effects of a magnetic
field are incorporated through  the Peierls phase as
\cite{Peierls1933}
\begin{equation}
t_{ij}\rightarrow t_{ij} \times \exp\left(\frac{ie}{\hbar} \int_{\mathbf{x}_i}^{\mathbf{x}_j}
d\mathbf{s} \mathbf{A}(\mathbf{x})\right)\,,
\end{equation}
where $\mathbf{x}_i$ and $\mathbf{x}_j$ are the positions of atom $i$ and $j$,
respectively, and $\mathbf{A}(\mathbf{x})$ is the magnetic vector potential.

The quantum dots are constructed by ``cutting'' the desired shape out
of the hexagonal graphene grid. For a shape that is smooth on the
length scale of the lattice constant as considered here, this results
in edges with a locally well-defined orientation (\emph{smooth edges},
see Fig.~\ref{sketch}(b)). In order to account for edge disorder on
the lattice scale (\emph{rough edges}, see Fig.~\ref{sketch}(c)), we
adopt the disorder model introduced in Ref.~\onlinecite{Mucciolo2009}:
Starting from the smooth edge, atoms at the boundary are removed
randomly with probability $p$, with dangling bonds removed after each
pass. This procedure is repeated $N_\text{sweep}$ times.

The energy spectrum of the dot tight-binding Hamiltonian is calculated
numerically using standard direct eigenvalue
algorithms\cite{Lapack} and matrix bandwidth reduction
techniques\cite{Gibbs1976} if a large part of the spectrum is needed.
In contrast, if only a few eigenvalues and -vectors are sought,
we apply an iterative technique\cite{ARPACK} in shift-and-invert
mode\cite{MUMPS}.

\subsection{Systems with electron-hole symmetry}

\begin{figure}
\includegraphics[width=\linewidth]{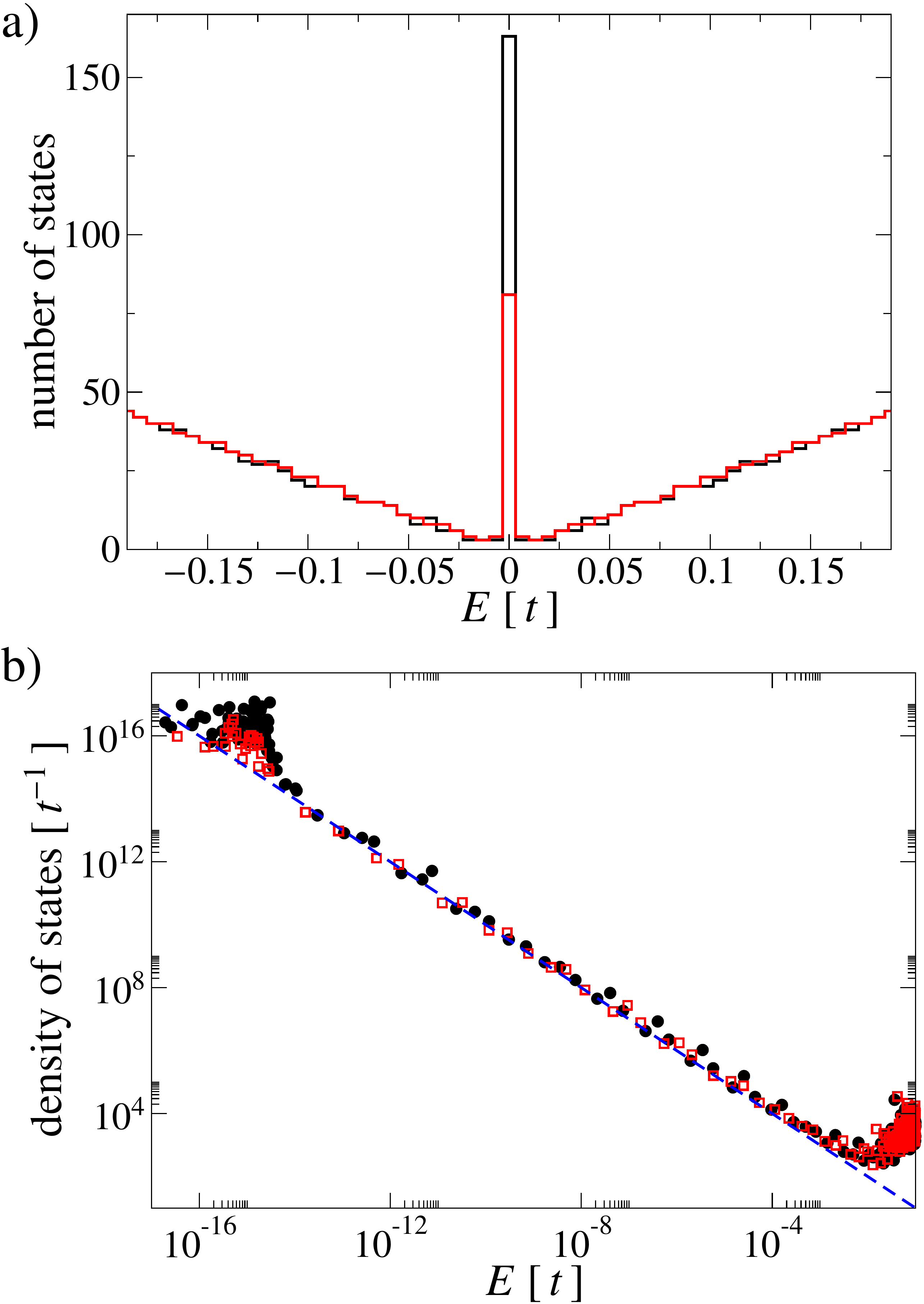}
\caption{Electronic states in a graphene quantum dot close to the
  Dirac point. The graphene quantum dot has the shape of a deformed
  circle (see Fig.~\ref{sketch} and footnote \onlinecite{dotshape}) with
  $R=160 a \approx 40\,\text{nm}$, and we consider both smooth and
  rough edges as shown in Fig.~\ref{sketch}(b) and (c) respectively.
  The parameters for the edge disorder are $N_\text{sweep}=5$ and $p=0.05$
  (see the main text for a discussion of the edge disorder model). (a)
  Number of states per energy interval $\Delta E$ for a quantum dot
  with smooth (black lines) and rough edges (red lines), with $\Delta
  E=0.4 t/61$. (b) Density of states estimated numerically from
  Eq.~\eqref{numdos} for a quantum dot with smooth (black symbols) and
  rough edges (red symbols). For comparison, the blue dashed line shows a
  $1/E$-dependence.}\label{fig:dot-tp0}
\end{figure}

We first focus on the electron-hole symmetric case, i.e.~$t'=0$ and
the absence of potentials. Fig.~\ref{fig:dot-tp0}(a) shows the
number of states $N(E)$ per energy interval $\Delta E$ for dots with
smooth and rough edges. We can clearly identify the edge states
close to $E=0$ and the linearly increasing bulk density of states.
Approximating the circumference of the dot as $L\approx 2\pi R$,
Eq.~\eqref{numedgestates1} predicts $N\approx 170$ edge states for a
quantum dot with a smooth edge, which is in very good agreement with
$N=169\pm 6$ edge states obtained from the numerical simulation by
summing over the three central bins, where the number of states
differs noticeably from the linear bulk density of states. The
number of edge states $N'$ in the dot with atomic scale disorder can
be estimated from Eq.~\eqref{numedgestates2} by approximating
$\alpha \approx p N_\text{sweep} $ yielding $N' \approx 0.5 N$ for
the disorder parameters used in the simulation ($N_\text{sweep}=5$,
$p=0.05$), again in good agreement with the numerical simulations.

In order to examine the behavior of the edge state density of states
in more detailed close to $E=0$, we estimate the density of states
numerically as
\begin{equation}\label{numdos}
\rho((E_{i+1}+E_{i})/2) = \frac{1}{E_{i+1} - E_i}\,,
\end{equation}
where $E_i$ is the energy of the $i$-th state in the dot.
Fig.~\ref{fig:dot-tp0}(b) shows the numerically
computed $\rho(E)$ for quantum dots with smooth and rough edges. As
predicted in Eq.~\eqref{oneoverE}, we find a $1/E$-dependence close
to $E=0$; quite remarkably, we find an excellent agreement with this
scaling for more than ten orders of magnitude. The clustering of data points at
$\rho(E)=10^{16}t^{-1}$ is due to the finite precision in the
numerical calculations. It should be noted that we found this
remarkable agreement with theoretical predictions without averaging
over an energy window or different dot shapes, implying that the
spectrum of edge states is highly non-random even in a quantum dot
with random shape. We come back to this point in Sec.~\ref{sec:level}.

\subsection{Broken electron-hole symmetry}

\begin{figure}
\includegraphics[width=\linewidth]{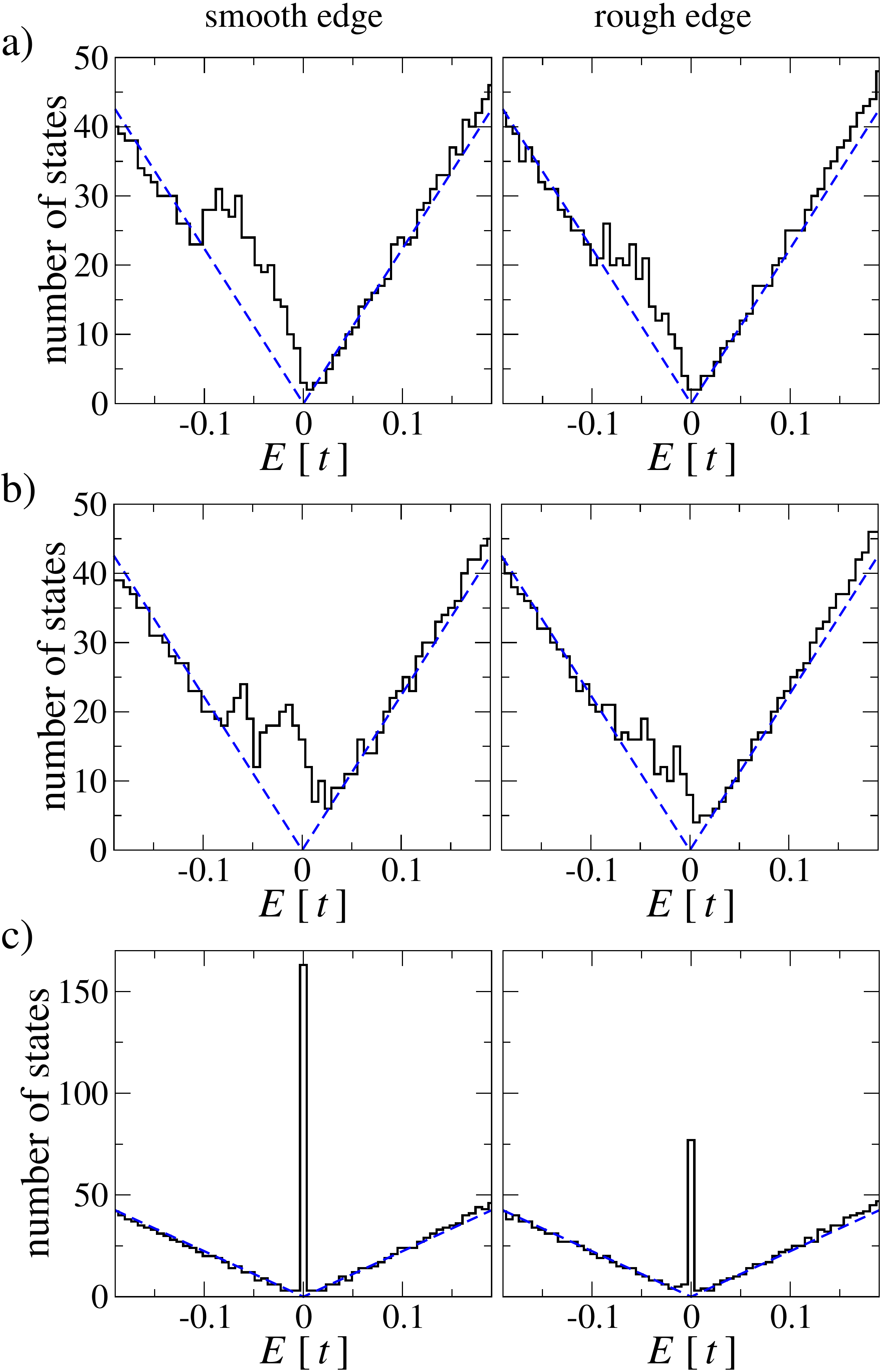}
\caption{Number of states (black lines) per energy interval $\Delta E$
  for a graphene quantum dot with smooth (left panels) and rough
  (right panels) edges. We show results for situations with broken
  electron-hole symmetry: (a) finite next-nearest neighbor hopping and
  no edge potential ($t'=0.1t$ and $U_0=0$) and (b,c) finite
  next-nearest neighbor hopping including an edge potential
  ($t'=0.1t$, with (b) $p_\text{edge}=0.25$ and $U_0=0.2t$, and (c)
  $p_\text{edge}=1$ and $U_0=0.1t$). The remaining parameters are as in
  Fig.~\ref{fig:dot-tp0}. The blue dashed lines show the number of
  bulk states $N_\text{bulk}$ estimated from the linear density of
  states of the Dirac dispersion
  Eq.~\eqref{diracdos}.}\label{fig:dot-tprime}
\end{figure}

Next we focus on perturbations breaking the electron-hole
symmetry. For this we consider a finite next-nearest neighbor hopping
$t'$ uniformly within the quantum dot, as well as a random potential at the
quantum dot edge, where an energy $U_0$ is assigned to edge atoms
with probability $p_\text{edge}$.

Fig.~\ref{fig:dot-tprime}(a) shows the number of states per energy window
$\Delta E$ for finite $t'$, but in the absence of an edge potential.
In order to identify the edge states properly, we compare the numerical
data including the edge states to the number of bulk states estimated from
the linear Dirac density of states,\cite{NGPNG09}
\begin{equation}\label{diracdos}
N_\text{bulk} (E) = \frac{2 \left|E\right| R^2}{v_\text{F}^2}
\end{equation}
approximating the area of the quantum dot as $A=\pi R^2$. The excess
edge state density of states can be clearly identified, both in the
case of smooth and rough edges. The bulk density of states close to
$E=0$ is unaffected by a finite $t'$, the effect of electron-hole
asymmetry on the bulk states only shows for energies $\left|E\right|>0.1t$.  The
central edge state peak observed for $t'=0$
(c.f.~Fig.\ref{fig:dot-tp0}) is broadened and shifted towards the
hole side, but the total number of edge states remains unchanged
from the $t'=0$ case. The excess density due to the edge states
is approximately constant in the energy range between $t'=-0.1t$ and
0, in accordance with the prediction from Eq.~\eqref{eq:edge_density}.
As before, atomic scale edge disorder only
results in a reduction of the total number of edge states.

The presence of an additional edge potential changes the
energy range of the edge states. In Fig.~\ref{fig:dot-tprime}(b)
we show results for an average edge potential $\Delta\epsilon=0.05t$.
Correspondingly, the majority of the edge states occupies uniformly
an energy window between $\Delta\epsilon-t'=-0.05t$ and $0$. A few states
can still be found beyond this energy window, as the randomness of
the edge potential has been neglected in the arguments of
Section \ref{sec:dispersion}. Instead, if the edge potential is
uniform, the dispersion of the edge state due to next-nearest neighbor
hopping can be cancelled exactly by $\Delta\epsilon=-t'$, as
shown in Fig.~\ref{fig:dot-tprime}(c). This particular example strikingly shows
the equivalence of next-nearest neighbor hopping and an edge potential,
as predicted in Ref.~\onlinecite{Sasaki2009}.

The narrowing of the energy band width occupied by the edge state due
to an edge potential may also be a possible explanation (amongst
others\cite{Sasaki2007b}) for the fact that STM measurements on zigzag
graphene edges found a peak in the density of states only a few tens
of meV below the Dirac point,\cite{Netal05b,Netal06}
far less than expected from estimated values of the
next-nearest neighbor hopping\cite{NGPNG09}.

\subsection{Broken time-reversal symmetry: Finite magnetic field}

\begin{figure}
\includegraphics[width=\linewidth]{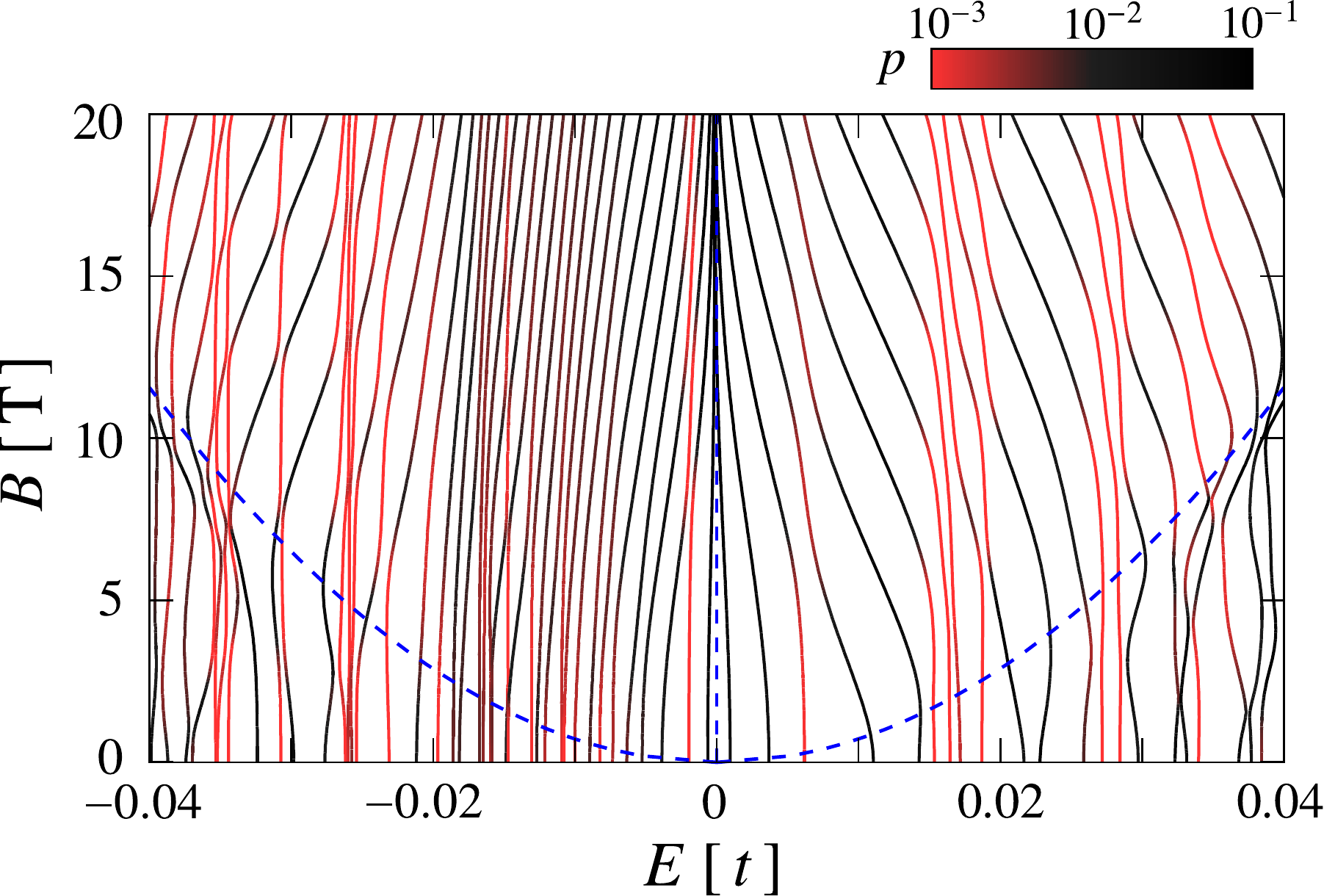}
\caption{Magnetic field dependence of the energy levels (black lines)
  in a desymmetrized quantum dot with $R=100a$ (deformed circle as
  shown in Fig.~\ref{sketch}, c.f.~footnote \onlinecite{dotshape}). The participation
  ratio $p$ of the states is color-encoded, with the most strongly localized states in red.
  The blue dashed lines indicate the energy of the $n=0,\pm1$ Landau levels of
  graphene. The calculations
  includes finite next-nearest neighbor hopping $t'=0.1t$ and
  a random edge potential with $p_\text{edge}=0.25$ and $U_0=0.2t$.}\label{fig:Bfield}
\end{figure}

We now consider the effects of a finite magnetic field on the edge
state energies. The evolution of edge states in a magnetic field has
been studied theoretically for special geometries and a particle-hole
symmetric spectrum\cite{Bahamon09,Kim10}.  Recently, the magnetic
field dependence of the energy levels in a graphene quantum dot has
been also been subject to an experimental investigation
\cite{Guttinger2009}.  However, in the theoretical calculations used
to interprete these experiments the graphene edge states were excluded
artificially. As we show below, the presence of edge states results in
a much richer magnetic field dependence of energy levels in a graphene
dot, in particular when particle-hole symmetry is broken.

In Fig.~\ref{fig:Bfield} we show the numerically calculated magnetic
field dependence of the energy levels in a graphene quantum dot close
to the Dirac point, for finite $t'$ and edge potential. In order to distinguish between
edge and bulk states, we also plot the participation ratio\cite{Bell1970,Bell1972}
\begin{equation}
p=\frac{\left(\sum_i \left|\psi(i)\right|^2\right)^2}{N \sum_i \left|\psi(i)\right|^4}
\end{equation}
where the index $i$ runs over atomic sites and $N$ denotes the total number of
atoms in the dot. The participation ratio $p$ can be interpreted as the fraction of atoms
occupied by an electron for a given energy level. Thus, $p\sim 1$ for extended states
($p\approx0.3-0.4$ in quantum dots) and $p\ll 1$ for localized edge states
($p\approx 10^{-4}-10^{-2}$).

Instead of a uniform flow of energy levels towards the
$n=$ Landau level as calculated in Ref.~\onlinecite{Guttinger2009}, we observe
that the most strongly localized states only show a very weak
magnetic field dependence (apart from avoided crossings), leading
to a far richer energy spectrum. Note that this effect is
most prominent on the hole side of the spectrum where the majority
of the edge states reside, as can be simply
seen by comparing the number of states for $E>0$ and $E<0$.
This weak magnetic field dependence of the localized edge states
can be understood from the fact that bulk states start to be affected by
the magnetic field when the cyclotron radius becomes comparable to the
dot size, whereas edge state energies are expected to only change
significantly when the cyclotron radius becomes comparable to the edge
state \emph{decay length} which is much smaller than the dot
dimensions.

Note that this type of behavior is similar to the magnetic field dependence
of the low-energy spectrum of graphene in the presence
of lattice vacancies.\cite{Libisch2010} In fact, such vacancies can be considered as internal
edges and also carry a localized state.

Hence, magnetic field independent energy levels are characteristic
for localized (edge) states. In the light of this observation, it would be very interesting to see if
experiments can identify such states, which would be a strong indication
for the presence of such states.

\subsection{Level statistics of edge states}\label{sec:level}

The bulk states of chaotic graphene quantum dots confined by lattice
termination have been shown to follow the level statistics of the
Gaussian orthogonal ensemble (GOE), as expected for a system with
time-reversal symmetry\cite{Wetal09,Libisch09} (scattering at the quantum dot
boundary mixes the $K$ and $K'$-valley). The edge states however are
tied to the boundary of the quantum dot only, and should not
necessarily follow the same level statistics as the extended
states. Instead, being localized states they are rather expected to
follow Poisson statistics, as has also been noted in
Ref.~\onlinecite{Wetal09}, but not been demonstrated explicitly.

To check these expectations we have studied the level spacing
distribution of edge states in quantum dots. For this purpose
we have identified edge states using the participation ratio and
worked with the edge state spectrum alone. This spectrum has been
unfolded\cite{Mehta} using the average density of states and scaled
to an average level spacing of unity. The
distribution $P(S)$ of the nearest-neighbor level spacings $S$ in the
unfolded spectrum is then normalized such that
$\int P(S) dS=1$ and $\int S P(S) dS=1$.

\begin{figure}
\includegraphics[width=\linewidth]{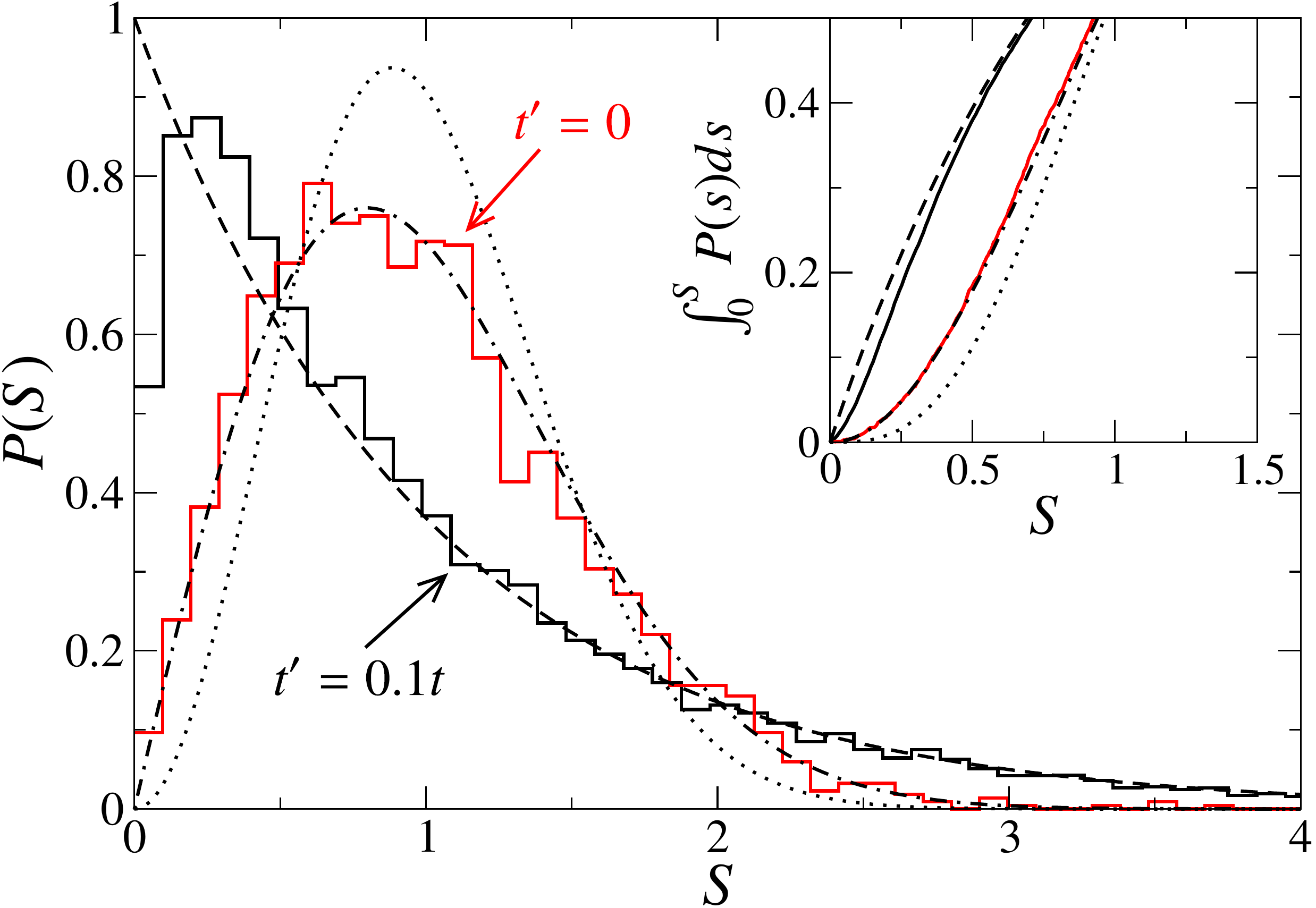}
\caption{Level spacing distributions for quantum dots with smooth edges
for $t'=0$ (solid red curve) and $t'=0.1t$ (solid black curve), together with the
theoretical predictions for Poisson statistics (dashed line), the
Gaussian orthogonal ensemble (dash-dotted line), and the Gaussian unitary ensemble
(dotted line). The inset shows details the integrated level spacing distribution
for small level spacings $S$ (same line colors and -types as the main plot).
The level distribution statistics has been obtained by averaging individual
level distributions from 100 quantum dots similar to the type
given in footnote \onlinecite{dotshape}, with average radius $R=160a$. A state
has been identified as an edge state, if its participation ratio
$p_i<0.05$.\cite{footnote_idedge}
For $t'=0$ we have also omitted all states with an energy smaller than the numerical
precision.
 }\label{levelstatistics}
\end{figure}

Fig.~\ref{levelstatistics} shows the level spacing distributions
for the electron-hole symmetric case ($t'=0$) and for
broken electron-hole symmetry ($t'=0.1t$). Surprisingly, the edge states
follow the GOE statistics if $t'=0$. Only if a finite $t'$ is included, they
exhibit a statistics close to Poisson. These classifications are
additionally corroborated by the integrated level spacing distributions
shown in the inset of Fig.~\ref{levelstatistics}.

\begin{figure}
\includegraphics[width=\linewidth]{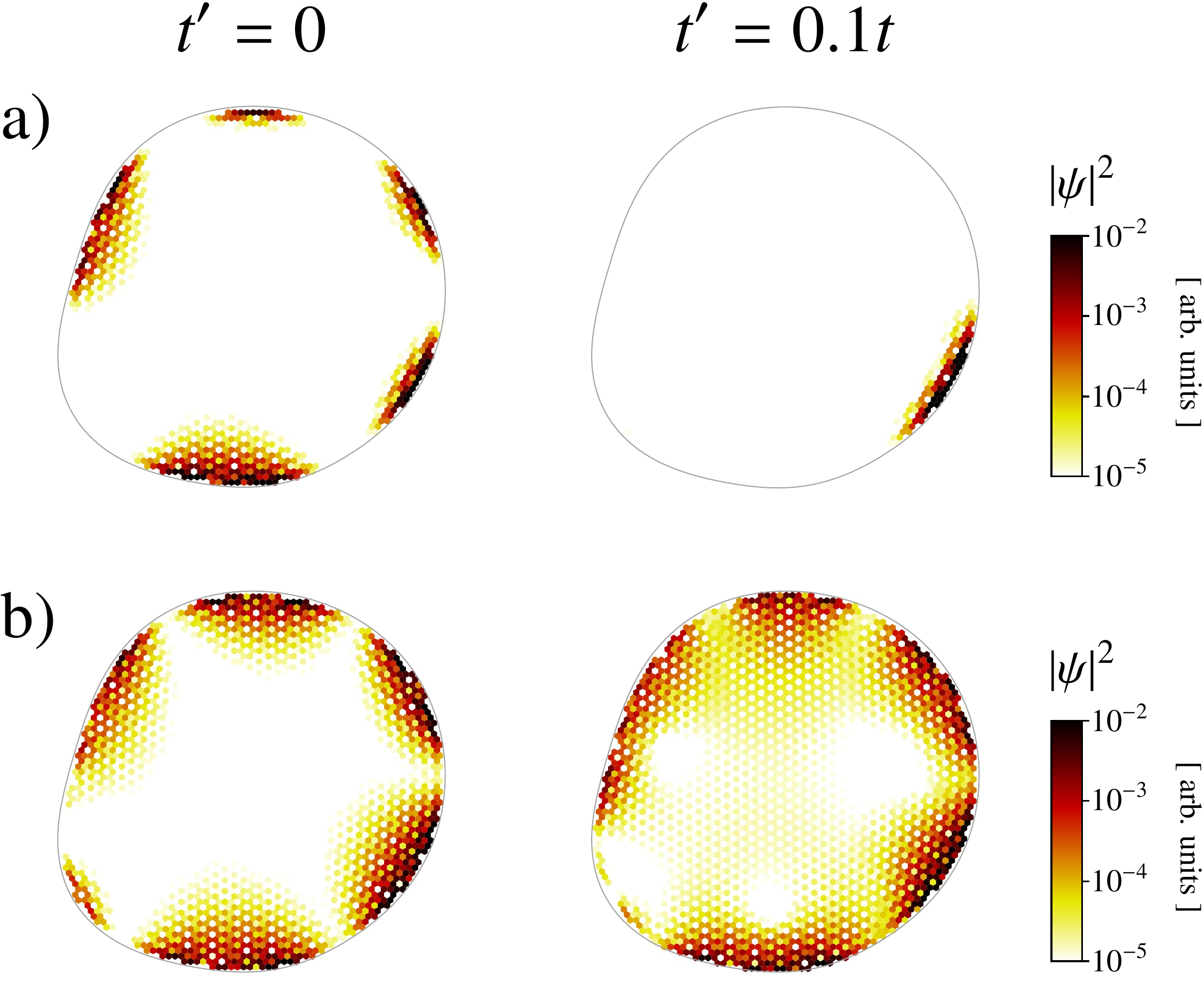}
\caption{Color plot\cite{footnote_colorplots} 
of wave function density in a graphene quantum dot (shape as
described in footnote \onlinecite{dotshape}) for the e-h symmetric case ($t'=0$, left column)
and for broken e-h symmetry ($t'=0.1t$, right column) on the examples of a mode that is (a) strongly decaying
and (b) slowly decaying into the bulk. Note that for presentation purposes we have chosen a rather small
dot ($R=30a$), but the behavior does not change qualitatively for larger dots.
 }\label{levelstatistics-wf}
\end{figure}

This striking difference in level statistics can be explained by the
different nature of the wave functions: The graphene Hamiltonian
exhibits a chiral symmetry for $t'=0$ that results in an equal
occupation probability of sublattice A and B for every individual wave
function.\cite{BF06} Since the edge wave function at a certain type of
zigzag edge is nonzero only on one sublattice, every eigenstate for
$t'=0$ must also occupy another part of the boundary of the opposite
kind, as illustrated in Fig.~\ref{levelstatistics-wf}.  This leads to an artificial
long-range coupling between edge states and thus to level repulsion,
resulting finally in GOE statistics. If this chiral symmetry is
broken, for example by next-nearest neighbor
hopping,\cite{footnote_oldpaper} edge state wave functions may be
localized at a single edge only
(Fig.~\ref{levelstatistics-wf}(a)).  Whereas edge states
localized at the same part of the boundary still may feel level
repulsion, parts that are further away may only interact via
hybridization with bulk states which typically happens for edges
states decaying further into the bulk, as seen in
Fig.~\ref{levelstatistics-wf}(b). For the type of quantum dots under
consideration (Fig.~\ref{sketch}), this results in six approximately
independent series of energy levels, and hence an approximate Poisson
statistics.

A finite next-nearest neighbor hopping $t'$ (or another chiral
symmetry breaking term) thus does not only change properties of the
edge states quantitatively, but leads to a striking, qualitatively
different level statistics.

\section{Discussion and physical implications} \label{sec_discussion}

\subsection{Formation of magnetic moments at the edges}
An extensively discussed topic in the graphene literature is the
formation of localized moments at
boundaries\cite{H01,HE02,SCL06,FP07,YK08,WSG08,WSSG08,RYL09}. The
previous analysis allows us to set approximate bounds on the maximum
magnetic moment in a graphene quantum dot.

The interaction energy between two electrons of opposite spin in a boundary state of area
$k_i^{-1} \times R\approx a\times R$  is:
\begin{align}
E_{ee} \approx \frac{e^2}{R} \log \left( \frac{R}{a}
\right) \label{Eee}
\end{align}
 where $e$ is the electronic
charge. States with energies $\left| \epsilon_i - E_F \right|
\lesssim E_{ee}^i$ will be spin polarized. Since the density of edge
states is nearly  constant and given by Eq.~\eqref{eq:edge_density},
the position of the Fermi level is not relevant.
Using the density of states given in
Eq.~\eqref{eq:edge_density}, we obtain for the number of spins in a
quantum dot:
\begin{multline}
N_\textrm{spins} \approx E_{ee} \rho_\textrm{edge} = c (1-2 \alpha)\frac{e^2}{a E_0}\log \left( \frac{R}{a} \right)\\
 \sim 20 (1-2 \alpha)\log \left( \frac{R}{a} \right)\label{numspins},
\end{multline}
where for last estimate we took $E_0=0.3$ eV. The maximal number of
polarized spins depends only logarithmically on the size of the dot.

In general, the states at the edge of a quantum dot will belong to one
of the two sublattices with equal probability. States localized at
different sublattices interact antiferromagnetically\cite{BFS07}. If
we neglect this interaction, we expect a maximum magnetic moment
comparable with $N_\textrm{spins}$. When the antiferromagnetic
interaction contributes to the formation of the total magnetic moment,
its value will be proportional to the number of uncompensated sites at
the edges, which will scale as $\sqrt{N_\textrm{spins}}$.

\subsection{Fraction of edge states}
Our results suggests that edge and bulk states can coexist in a range of energy of order $E_0$ near the Dirac point. From Eqs.~\eqref{eq:edge_density} and~\eqref{diracdos}, the average ratio between edge and bulk states in this energy range is
\begin{align}
\frac{N_{edge}}{N_{bulk}} &\approx c ( 1 - 2 \alpha ) \frac{v_\text{F}^2}{E_0^2 a R}
\end{align}
This gives, for a diameter of 100nm and $E_0 = 0.3$eV an upper bound of $N_{edge} / N_{bulk} \lesssim 1/2$.

\subsection{Detection in antidot lattices}
A conclusive way of detecting the existence of edge states can be the
measurement of their contribution to the electronic
compressibility. It is hard to detect the edge states in a single
quantum dot because the ground state properties are dominated by the
charging energy. Also, the contribution of edge states to the density
of states in most large-scale samples will be negligible compared to
the bulk contribution. However it is possible to circumvent both
problems in antidot lattices. The Coulomb energy does not play a role
in this case due to absence of confinement. On the other hand, the
existence of multiple antidots allows us to reach a large edge-area
ratio. To estimate whether it is possible to detect edge states, we
use the value of minimal compressibility (or the minimal density of
states) of bulk graphene Ref.~\onlinecite{Martin2008}
\begin{align}
\frac{\partial\mu}{\partial n} &=3 \times 10^{-10}
\textrm{meV}\textrm{cm}^{2} \label{exp}
\end{align}
and we assume that the width of the band of edge states is around
$E_0 \approx 0.3$ eV.

We consider an antidot lattice with antidot size $L$ of the same
order of magnitude as the antidot spacing. Using the analysis in the
previous section, the density of states per unit area associated to
the edge states is:
\begin{align}
N_{area}^{-1} ( E ) &\approx E_0 a L
\end{align}
Comparing this expression with eq.~\ref{exp}, and using $E_0 \approx
0.3$ eV, we find that the contribution from the edge states is
comparable to the bulk inverse compressibility for $L \lesssim 1
\mu$m. Hence, the additional density of states near the edge will be
visible in compressibility measurements using a single electron
transistor (SET) since the size of the SET tip is around 100
nanometers.\cite{Martin2008} Our results may be the reason
of \emph{p}-doping observed in antidot lattices experimentally.\cite{Lee08,Hey10}

\section{Conclusions}
\label{Sec_conclusion}
We have analyzed generic properties of the electronic spectrum in
graphene quantum dots. We find that some of the electronic states will
be localized at the edges and form a narrow band. The density of
states in this band is $\propto 1/E$ in graphene dot without
electron-hole symmetry breaking perturbations. In presence of such
perturbations, the density of the edge states is approximately
constant and scales as $R / a E_0$, where $R$ is the dot radius, $a$
is the lattice constant, and $E_0$ is an energy scale which describes
the edge potentials and next-nearest neighbor hopping.

If chiral symmetry is present, the edge states experience strong level
repulsion and are described by the Gaussian orthogonal ensemble. Chiral
symmetry breaking terms (such as next-nearest neighbor hopping) however
lift this spurious level repulsion leading to the Poissonian statistics expected
for localized states. In contrast, extended states will be
described by the orthogonal or unitary ensembles, depending on the
strength of the intervalley scattering at the
boundaries\cite{RK08,Wetal09}.

Having an analytical model for the edge states allows us to estimate
the maximum spin polarization due to the presence of edge states. We
predict that the additional density of states due to edge states will be
visible in SET experiments. Effect of edge states on transport in
quantum dots and more detailed investigation of interaction effects
remains a direction for further research.

\section{Acknowledgements}

We are grateful for useful discussions to C.~W.~J. Beenakker, K. Ensslin, A.~K. Geim, I.~V. Grigorieva, K.~S. Novoselov, J.~H. Smet,  and C. Stampfer. FG is supported by
MEC (Spain) Grants FIS2008-00124 and CONSOLIDER CSD2007-00010, and
also the Comunidad de Madrid, through CITECNOMIK. MW is supported by the
Deutscher Akademischer Austausch Dienst DAAD. AA is supported by the Dutch Science Foundation NWO/FOM and by the Eurocores
program EuroGraphene.


\begin{thebibliography}{10}%
\makeatletter
\providecommand \@ifxundefined [1]{%
 \ifx #1\undefined \expandafter \@firstoftwo
 \else \expandafter \@secondoftwo
\fi
}%
\providecommand \@ifnum [1]{%
 \ifnum #1\expandafter \@firstoftwo
 \else \expandafter \@secondoftwo
\fi
}%
\providecommand \enquote [1]{``#1''}%
\providecommand \bibnamefont  [1]{#1}%
\providecommand \bibfnamefont [1]{#1}%
\providecommand \citenamefont [1]{#1}%
\providecommand\href[0]{\@sanitize\@href}%
\providecommand\@href[1]{\endgroup\@@startlink{#1}\endgroup\@@href}%
\providecommand\@@href[1]{#1\@@endlink}%
\providecommand \@sanitize [0]{\begingroup\catcode`\&12\catcode`\#12\relax}%
\@ifxundefined \pdfoutput {\@firstoftwo}{%
 \@ifnum{\z@=\pdfoutput}{\@firstoftwo}{\@secondoftwo}%
}{%
 \providecommand\@@startlink[1]{\leavevmode}%
 \providecommand\@@endlink[0]{}%
}{%
 \providecommand\@@startlink[1]{%
  \leavevmode
  \pdfstartlink
   attr{/Border[0 0 1 ]/H/I/C[0 1 1]}%
   user{/Subtype/Link/A<</Type/Action/S/URI/URI(#1)>>}%
  \relax
 }%
 \providecommand\@@endlink[0]{\pdfendlink}%
}%
\providecommand \url  [0]{\begingroup\@sanitize \@url }%
\providecommand \@url [1]{\endgroup\@href {#1}{\urlprefix}}%
\providecommand \urlprefix [0]{URL }%
\providecommand \Eprint[0]{\href }%
\@ifxundefined \urlstyle {%
  \providecommand \doi [1]{doi:\discretionary{}{}{}#1}%
}{%
  \providecommand \doi [0]{doi:\discretionary{}{}{}\begingroup
  \urlstyle{rm}\Url }%
}%
\providecommand \doibase [0]{http://dx.doi.org/}%
\providecommand \Doi[1]{\href{\doibase#1}}%
\providecommand \bibAnnote [3]{%
  \BibitemShut{#1}%
  \begin{quotation}\noindent
    \textsc{Key:}\ #2\\\textsc{Annotation:}\ #3%
  \end{quotation}%
}%
\providecommand \bibAnnoteFile [2]{%
  \IfFileExists{#2}{\bibAnnote {#1} {#2} {\input{#2}}}{}%
}%
\providecommand \typeout [0]{\immediate \write \m@ne }%
\providecommand \selectlanguage [0]{\@gobble}%
\providecommand \bibinfo [0]{\@secondoftwo}%
\providecommand \bibfield [0]{\@secondoftwo}%
\providecommand \translation [1]{[#1]}%
\providecommand \BibitemOpen[0]{}%
\providecommand \bibitemStop [0]{}%
\providecommand \bibitemNoStop [0]{.\EOS\space}%
\providecommand \EOS [0]{\spacefactor3000\relax}%
\providecommand \BibitemShut [1]{\csname bibitem#1\endcsname}%
\bibitem{Netal04}%
  \BibitemOpen
  \bibfield{author}{%
  \bibinfo {author} {\bibfnamefont{K.~S.}\ \bibnamefont{Novoselov}}, \bibinfo
  {author} {\bibfnamefont{A.~K.}\ \bibnamefont{Geim}}, \bibinfo {author}
  {\bibfnamefont{S.~V.}\ \bibnamefont{Morozov}}, \bibinfo {author}
  {\bibfnamefont{D.}~\bibnamefont{Jiang}}, \bibinfo {author}
  {\bibfnamefont{Y.}~\bibnamefont{Zhang}}, \bibinfo {author}
  {\bibfnamefont{S.~V.}\ \bibnamefont{Dubonos}}, \bibinfo {author}
  {\bibfnamefont{I.~V.}\ \bibnamefont{Grigorieva}},\ and\ \bibinfo {author}
  {\bibfnamefont{A.~A.}\ \bibnamefont{Firsov}},\ }%
  \bibfield{journal}{%
  \bibinfo {journal} {Science}\ }%
  \textbf{\bibinfo {volume} {306}},\ \bibinfo {pages} {666} (\bibinfo {year}
  {2004})%
  \bibAnnoteFile{NoStop}{Netal04}%
\bibitem{Netal05}%
  \BibitemOpen
  \bibfield{author}{%
  \bibinfo {author} {\bibfnamefont{K.~S.}\ \bibnamefont{Novoselov}}, \bibinfo
  {author} {\bibfnamefont{D.}~\bibnamefont{Jiang}}, \bibinfo {author}
  {\bibfnamefont{F.}~\bibnamefont{Schedin}}, \bibinfo {author}
  {\bibfnamefont{T.~J.}\ \bibnamefont{Booth}}, \bibinfo {author}
  {\bibfnamefont{V.~V.}\ \bibnamefont{Khotkevich}}, \bibinfo {author}
  {\bibfnamefont{S.~V.}\ \bibnamefont{Morozov}},\ and\ \bibinfo {author}
  {\bibfnamefont{A.~K.}\ \bibnamefont{Geim}},\ }%
  \bibfield{journal}{%
  \bibinfo {journal} {Proc. Natl. Acad. Sci. U.S.A.}\ }%
  \textbf{\bibinfo {volume} {102}},\ \bibinfo {pages} {10451} (\bibinfo {year}
  {2005})%
  \bibAnnoteFile{NoStop}{Netal05}%
\bibitem{GN07}%
  \BibitemOpen
  \bibfield{author}{%
  \bibinfo {author} {\bibfnamefont{A.~K.}\ \bibnamefont{Geim}}\ and\ \bibinfo
  {author} {\bibfnamefont{K.~S.}\ \bibnamefont{Novoselov}},\ }%
  \bibfield{journal}{%
  \bibinfo {journal} {Nature Materials}\ }%
  \textbf{\bibinfo {volume} {6}},\ \bibinfo {pages} {183} (\bibinfo {year}
  {2007})%
  \bibAnnoteFile{NoStop}{GN07}%
\bibitem{ACP07}%
  \BibitemOpen
  \bibfield{author}{%
  \bibinfo {author} {\bibfnamefont{P.}~\bibnamefont{Avouris}}, \bibinfo
  {author} {\bibfnamefont{Z.}~\bibnamefont{Chen}},\ and\ \bibinfo {author}
  {\bibfnamefont{V.}~\bibnamefont{Perebeinos}},\ }%
  \bibfield{journal}{%
  \bibinfo {journal} {Nature Nanotechnology}\ }%
  \textbf{\bibinfo {volume} {2}},\ \bibinfo {pages} {605} (\bibinfo {year}
  {2007})%
  \bibAnnoteFile{NoStop}{ACP07}%
\bibitem{NGPNG09}%
  \BibitemOpen
  \bibfield{author}{%
  \bibinfo {author} {\bibfnamefont{A.~H.}\ \bibnamefont{{Castro Neto}}},
  \bibinfo {author} {\bibfnamefont{F.}~\bibnamefont{Guinea}}, \bibinfo {author}
  {\bibfnamefont{N.~M.~R.}\ \bibnamefont{Peres}}, \bibinfo {author}
  {\bibfnamefont{K.~S.}\ \bibnamefont{Novoselov}},\ and\ \bibinfo {author}
  {\bibfnamefont{A.~K.}\ \bibnamefont{Geim}},\ }%
  \bibfield{journal}{%
  \bibinfo {journal} {Rev. Mod. Phys.}\ }%
  \textbf{\bibinfo {volume} {81}},\ \bibinfo {pages} {109} (\bibinfo {year}
  {2009})%
  \bibAnnoteFile{NoStop}{NGPNG09}%
\bibitem{K94}%
  \BibitemOpen
  \bibfield{author}{%
  \bibinfo {author} {\bibfnamefont{D.~J.}\ \bibnamefont{Klein}},\ }%
  \bibfield{journal}{%
  \bibinfo {journal} {Chem. Phys. Lett.}\ }%
  \textbf{\bibinfo {volume} {217}},\ \bibinfo {pages} {261} (\bibinfo {year}
  {1994})%
  \bibAnnoteFile{NoStop}{K94}%
\bibitem{FWNK96}%
  \BibitemOpen
  \bibfield{author}{%
  \bibinfo {author} {\bibfnamefont{M.}~\bibnamefont{Fujita}}, \bibinfo {author}
  {\bibfnamefont{K.}~\bibnamefont{Wakabayashi}}, \bibinfo {author}
  {\bibfnamefont{K.}~\bibnamefont{Nakada}},\ and\ \bibinfo {author}
  {\bibfnamefont{K.}~\bibnamefont{Kusakabe}},\ }%
  \bibfield{journal}{%
  \bibinfo {journal} {J. Phys. Soc. Jpn.}\ }%
  \textbf{\bibinfo {volume} {65}},\ \bibinfo {pages} {1920} (\bibinfo {year}
  {1996})%
  \bibAnnoteFile{NoStop}{FWNK96}%
\bibitem{NFDD96}%
  \BibitemOpen
  \bibfield{author}{%
  \bibinfo {author} {\bibfnamefont{K.}~\bibnamefont{Nakada}}, \bibinfo {author}
  {\bibfnamefont{M.}~\bibnamefont{Fujita}}, \bibinfo {author}
  {\bibfnamefont{G.}~\bibnamefont{Dresselhaus}},\ and\ \bibinfo {author}
  {\bibfnamefont{M.~S.}\ \bibnamefont{Dresselhaus}},\ }%
  \bibfield{journal}{%
  \bibinfo {journal} {Phys. Rev. B}\ }%
  \textbf{\bibinfo {volume} {54}},\ \bibinfo {pages} {17954} (\bibinfo {year}
  {1996})%
  \bibAnnoteFile{NoStop}{NFDD96}%
\bibitem{AB08}%
  \BibitemOpen
  \bibfield{author}{%
  \bibinfo {author} {\bibfnamefont{A.~R.}\ \bibnamefont{Akhmerov}}\ and\
  \bibinfo {author} {\bibfnamefont{C.~W.~J.}\ \bibnamefont{Beenakker}},\ }%
  \bibfield{journal}{%
  \bibinfo {journal} {Phys. Rev. B}\ }%
  \textbf{\bibinfo {volume} {77}},\ \bibinfo {pages} {085423} (\bibinfo {year}
  {2008})%
  \bibAnnoteFile{NoStop}{AB08}%
\bibitem{CPLNG08}%
  \BibitemOpen
  \bibfield{author}{%
  \bibinfo {author} {\bibfnamefont{E.~V.}\ \bibnamefont{Castro}}, \bibinfo
  {author} {\bibnamefont{{N. M. R. Peres}}}, \bibinfo {author}
  {\bibnamefont{{J. M. B. Lopes dos Santos}}}, \bibinfo {author}
  {\bibnamefont{{A. H. Castro Neto}}},\ and\ \bibinfo {author}
  {\bibfnamefont{F.}~\bibnamefont{Guinea}},\ }%
  \bibfield{journal}{%
  \bibinfo {journal} {Phys. Rev. Lett.}\ }%
  \textbf{\bibinfo {volume} {100}},\ \bibinfo {pages} {026802} (\bibinfo {year}
  {2008})%
  \bibAnnoteFile{NoStop}{CPLNG08}%
\bibitem{SMMB08}%
  \BibitemOpen
  \bibfield{author}{%
  \bibinfo {author} {\bibfnamefont{B.}~\bibnamefont{Sahu}}, \bibinfo {author}
  {\bibfnamefont{H.}~\bibnamefont{Min}}, \bibinfo {author}
  {\bibfnamefont{A.~H.}\ \bibnamefont{MacDonald}},\ and\ \bibinfo {author}
  {\bibfnamefont{S.~K.}\ \bibnamefont{Banerjee}},\ }%
  \bibfield{journal}{%
  \bibinfo {journal} {Phys. Rev. B}\ }%
  \textbf{\bibinfo {volume} {78}},\ \bibinfo {pages} {045404} (\bibinfo {year}
  {2008})%
  \bibAnnoteFile{NoStop}{SMMB08}%
\bibitem{EVCastro2008}%
  \BibitemOpen
  \bibfield{author}{%
  \bibinfo {author} {\bibfnamefont{E.~V.}\ \bibnamefont{Castro}}, \bibinfo
  {author} {\bibfnamefont{N.~M.~R.}\ \bibnamefont{Peres}},\ and\ \bibinfo
  {author} {\bibfnamefont{J.~M.~B.}\ \bibnamefont{Lopes~dos Santos}},\ }%
  \bibfield{journal}{%
  \bibinfo {journal} {Europhys. Lett.}\ }%
  \textbf{\bibinfo {volume} {84}},\ \bibinfo {pages} {17001} (\bibinfo {year}
  {2008})%
  \bibAnnoteFile{NoStop}{EVCastro2008}%
\bibitem{Netal05b}%
  \BibitemOpen
  \bibfield{author}{%
  \bibinfo {author} {\bibfnamefont{Y.}~\bibnamefont{Niimi}}, \bibinfo {author}
  {\bibfnamefont{T.}~\bibnamefont{Matsui}}, \bibinfo {author}
  {\bibfnamefont{H.}~\bibnamefont{Kambara}}, \bibinfo {author}
  {\bibfnamefont{K.}~\bibnamefont{Tagami}}, \bibinfo {author}
  {\bibfnamefont{M.}~\bibnamefont{Tsukada}},\ and\ \bibinfo {author}
  {\bibfnamefont{H.}~\bibnamefont{Fukuyama}},\ }%
  \bibfield{journal}{%
  \bibinfo {journal} {Appl. Surf. Sci.}\ }%
  \textbf{\bibinfo {volume} {241}},\ \bibinfo {pages} {43} (\bibinfo {year}
  {2005})%
  \bibAnnoteFile{NoStop}{Netal05b}%
\bibitem{KFEKK05}%
  \BibitemOpen
  \bibfield{author}{%
  \bibinfo {author} {\bibfnamefont{Y.}~\bibnamefont{Kobayashi}}, \bibinfo
  {author} {\bibfnamefont{K.-I.}\ \bibnamefont{Fukui}}, \bibinfo {author}
  {\bibfnamefont{T.}~\bibnamefont{Enoki}}, \bibinfo {author}
  {\bibfnamefont{K.}~\bibnamefont{Kusakabe}},\ and\ \bibinfo {author}
  {\bibfnamefont{Y.}~\bibnamefont{Kaburagi}},\ }%
  \bibfield{journal}{%
  \bibinfo {journal} {Phys. Rev. B}\ }%
  \textbf{\bibinfo {volume} {71}},\ \bibinfo {pages} {193406} (\bibinfo {year}
  {2005})%
  \bibAnnoteFile{NoStop}{KFEKK05}%
\bibitem{Netal06}%
  \BibitemOpen
  \bibfield{author}{%
  \bibinfo {author} {\bibfnamefont{Y.}~\bibnamefont{Niimi}}, \bibinfo {author}
  {\bibfnamefont{T.}~\bibnamefont{Matsui}}, \bibinfo {author}
  {\bibfnamefont{H.}~\bibnamefont{Kambara}}, \bibinfo {author}
  {\bibfnamefont{K.}~\bibnamefont{Tagami}}, \bibinfo {author}
  {\bibfnamefont{M.}~\bibnamefont{Tsukada}},\ and\ \bibinfo {author}
  {\bibfnamefont{H.}~\bibnamefont{Fukuyama}},\ }%
  \bibfield{journal}{%
  \bibinfo {journal} {Phys. Rev. B}\ }%
  \textbf{\bibinfo {volume} {73}},\ \bibinfo {pages} {085421} (\bibinfo {year}
  {2006})%
  \bibAnnoteFile{NoStop}{Netal06}%
\bibitem{NIF98}%
  \BibitemOpen
  \bibfield{author}{%
  \bibinfo {author} {\bibfnamefont{K.}~\bibnamefont{Nakada}}, \bibinfo {author}
  {\bibfnamefont{M.}~\bibnamefont{Igami}},\ and\ \bibinfo {author}
  {\bibfnamefont{M.}~\bibnamefont{Fujita}},\ }%
  \bibfield{journal}{%
  \bibinfo {journal} {J. Phys. Soc. Jpn.}\ }%
  \textbf{\bibinfo {volume} {67}},\ \bibinfo {pages} {2388} (\bibinfo {year}
  {1998})%
  \bibAnnoteFile{NoStop}{NIF98}%
\bibitem{SCL06}%
  \BibitemOpen
  \bibfield{author}{%
  \bibinfo {author} {\bibfnamefont{Y.-W.}\ \bibnamefont{Son}}, \bibinfo
  {author} {\bibfnamefont{M.~L.}\ \bibnamefont{Cohen}},\ and\ \bibinfo {author}
  {\bibfnamefont{S.~G.}\ \bibnamefont{Louie}},\ }%
  \bibfield{journal}{%
  \bibinfo {journal} {Nature}\ }%
  \textbf{\bibinfo {volume} {444}},\ \bibinfo {pages} {347} (\bibinfo {year}
  {2006})%
  \bibAnnoteFile{NoStop}{SCL06}%
\bibitem{FP07}%
  \BibitemOpen
  \bibfield{author}{%
  \bibinfo {author} {\bibfnamefont{J.}~\bibnamefont{Fernandez-Rossier}}\ and\
  \bibinfo {author} {\bibfnamefont{J.~J.}\ \bibnamefont{Palacios}},\ }%
  \bibfield{journal}{%
  \bibinfo {journal} {Phys. Rev. Lett.}\ }%
  \textbf{\bibinfo {volume} {99}},\ \bibinfo {pages} {177204} (\bibinfo {year}
  {2007})%
  \bibAnnoteFile{NoStop}{FP07}%
\bibitem{Ezawa07}%
  \BibitemOpen
  \bibfield{author}{%
  \bibinfo {author} {\bibfnamefont{M.}~\bibnamefont{Ezawa}},\ }%
  \bibfield{journal}{%
  \bibinfo {journal} {Phys. Rev. B}\ }%
  \textbf{\bibinfo {volume} {76}},\ \bibinfo {pages} {245415} (\bibinfo {year}
  {2007})%
  \bibAnnoteFile{NoStop}{Ezawa07}%
\bibitem{YK08}%
  \BibitemOpen
  \bibfield{author}{%
  \bibinfo {author} {\bibfnamefont{O.~V.}\ \bibnamefont{Yazyev}}\ and\ \bibinfo
  {author} {\bibfnamefont{M.~I.}\ \bibnamefont{Katsnelson}},\ }%
  \bibfield{journal}{%
  \bibinfo {journal} {Phys. Rev. Lett.}\ }%
  \textbf{\bibinfo {volume} {100}},\ \bibinfo {pages} {047209} (\bibinfo {year}
  {2008})%
  \bibAnnoteFile{NoStop}{YK08}%
\bibitem{Wimmer2008}%
  \BibitemOpen
  \bibfield{author}{%
  \bibinfo {author} {\bibfnamefont{M.}~\bibnamefont{Wimmer}}, \bibinfo {author}
  {\bibfnamefont{I.}~\bibnamefont{Adagideli}}, \bibinfo {author}
  {\bibfnamefont{S.}~\bibnamefont{Berber}}, \bibinfo {author}
  {\bibfnamefont{D.}~\bibnamefont{Tom\'anek}},\ and\ \bibinfo {author}
  {\bibfnamefont{K.}~\bibnamefont{Richter}},\ }%
  \bibfield{journal}{%
  \Doi{10.1103/PhysRevLett.100.177207}{\bibinfo {journal} {Phys. Rev. Lett.}}\
  }%
  \textbf{\bibinfo {volume} {100}},\ \bibinfo {pages} {177207} (\bibinfo {year}
  {2008})%
  \bibAnnoteFile{NoStop}{Wimmer2008}%
\bibitem{WSG08}%
  \BibitemOpen
  \bibfield{author}{%
  \bibinfo {author} {\bibfnamefont{B.}~\bibnamefont{Wunsch}}, \bibinfo {author}
  {\bibfnamefont{T.}~\bibnamefont{Stauber}},\ and\ \bibinfo {author}
  {\bibfnamefont{F.}~\bibnamefont{Guinea}},\ }%
  \bibfield{journal}{%
  \bibinfo {journal} {Phys. Rev. B}\ }%
  \textbf{\bibinfo {volume} {77}},\ \bibinfo {pages} {035316} (\bibinfo {year}
  {2008})%
  \bibAnnoteFile{NoStop}{WSG08}%
\bibitem{WSSG08}%
  \BibitemOpen
  \bibfield{author}{%
  \bibinfo {author} {\bibfnamefont{B.}~\bibnamefont{Wunsch}}, \bibinfo {author}
  {\bibfnamefont{T.}~\bibnamefont{Stauber}}, \bibinfo {author}
  {\bibfnamefont{F.}~\bibnamefont{Sols}},\ and\ \bibinfo {author}
  {\bibfnamefont{F.}~\bibnamefont{Guinea}},\ }%
  \bibfield{journal}{%
  \bibinfo {journal} {Phys. Rev. Lett.}\ }%
  \textbf{\bibinfo {volume} {101}},\ \bibinfo {pages} {036803} (\bibinfo {year}
  {2008})%
  \bibAnnoteFile{NoStop}{WSSG08}%
\bibitem{RYL09}%
  \BibitemOpen
  \bibfield{author}{%
  \bibinfo {author} {\bibfnamefont{I.}~\bibnamefont{Romanovsky}}, \bibinfo
  {author} {\bibfnamefont{C.}~\bibnamefont{Yannouleas}},\ and\ \bibinfo
  {author} {\bibfnamefont{U.}~\bibnamefont{Landman}},\ }%
  \bibfield{journal}{%
  \bibinfo {journal} {Phys. Rev. B}\ }%
  \textbf{\bibinfo {volume} {79}},\ \bibinfo {pages} {075311} (\bibinfo {year}
  {2009})%
  \bibAnnoteFile{NoStop}{RYL09}%
\bibitem{Petal08}%
  \BibitemOpen
  \bibfield{author}{%
  \bibinfo {author} {\bibfnamefont{L.~A.}\ \bibnamefont{Ponomarenko}}, \bibinfo
  {author} {\bibfnamefont{F.}~\bibnamefont{Schedin}}, \bibinfo {author}
  {\bibfnamefont{M.~I.}\ \bibnamefont{Katsnelson}}, \bibinfo {author}
  {\bibfnamefont{R.}~\bibnamefont{Yang}}, \bibinfo {author}
  {\bibfnamefont{E.~W.}\ \bibnamefont{Hill}}, \bibinfo {author}
  {\bibfnamefont{K.~S.}\ \bibnamefont{Novoselov}},\ and\ \bibinfo {author}
  {\bibfnamefont{A.~K.}\ \bibnamefont{Geim}},\ }%
  \bibfield{journal}{%
  \bibinfo {journal} {Science}\ }%
  \textbf{\bibinfo {volume} {320}},\ \bibinfo {pages} {356} (\bibinfo {year}
  {2008})%
  \bibAnnoteFile{NoStop}{Petal08}%
\bibitem{Setal08}%
  \BibitemOpen
  \bibfield{author}{%
  \bibinfo {author} {\bibfnamefont{J.}~\bibnamefont{Guettinger}}, \bibinfo
  {author} {\bibfnamefont{C.}~\bibnamefont{Stampfer}}, \bibinfo {author}
  {\bibfnamefont{S.}~\bibnamefont{Hellmueller}}, \bibinfo {author}
  {\bibfnamefont{F.}~\bibnamefont{Molitor}}, \bibinfo {author}
  {\bibfnamefont{T.}~\bibnamefont{Ihn}},\ and\ \bibinfo {author}
  {\bibfnamefont{K.}~\bibnamefont{Ensslin}},\ }%
  \bibfield{journal}{%
  \bibinfo {journal} {Appl. Phys. Lett.}\ }%
  \textbf{\bibinfo {volume} {93}},\ \bibinfo {pages} {212102} (\bibinfo {year}
  {2008})%
  \bibAnnoteFile{NoStop}{Setal08}%
\bibitem{Stampfer2008}%
  \BibitemOpen
  \bibfield{author}{%
  \bibinfo {author} {\bibfnamefont{C.}~\bibnamefont{Stampfer}}, \bibinfo
  {author} {\bibfnamefont{J.}~\bibnamefont{G\"{u}ttinger}}, \bibinfo {author}
  {\bibfnamefont{F.}~\bibnamefont{Molitor}}, \bibinfo {author}
  {\bibfnamefont{D.}~\bibnamefont{Graf}}, \bibinfo {author}
  {\bibfnamefont{T.}~\bibnamefont{Ihn}},\ and\ \bibinfo {author}
  {\bibfnamefont{K.}~\bibnamefont{Ensslin}},\ }%
  \bibfield{journal}{%
  \bibinfo {journal} {Applied Physics Letters}\ }%
  \textbf{\bibinfo {volume} {92}},\ \bibinfo {eid} {012102} (\bibinfo {year}
  {2008})%
  \bibAnnoteFile{NoStop}{Stampfer2008}%
\bibitem{Guttinger2009}%
  \BibitemOpen
  \bibfield{author}{%
  \bibinfo {author} {\bibfnamefont{J.}~\bibnamefont{G\"{u}ttinger}}, \bibinfo
  {author} {\bibfnamefont{C.}~\bibnamefont{Stampfer}}, \bibinfo {author}
  {\bibfnamefont{F.}~\bibnamefont{Libisch}}, \bibinfo {author}
  {\bibfnamefont{T.}~\bibnamefont{Frey}}, \bibinfo {author}
  {\bibfnamefont{J.}~\bibnamefont{Burgd\"{o}rfer}}, \bibinfo {author}
  {\bibfnamefont{T.}~\bibnamefont{Ihn}},\ and\ \bibinfo {author}
  {\bibfnamefont{K.}~\bibnamefont{Ensslin}},\ }%
  \bibfield{journal}{%
  \Doi{10.1103/PhysRevLett.103.046810}{\bibinfo {journal} {Phys. Rev. Lett.}}\
  }%
  \textbf{\bibinfo {volume} {103}},\ \bibinfo {eid} {046810} (\bibinfo {year}
  {2009})%
  \bibAnnoteFile{NoStop}{Guttinger2009}%
\bibitem{Setal09b}%
  \BibitemOpen
  \bibfield{author}{%
  \bibinfo {author} {\bibfnamefont{S.}~\bibnamefont{Schnez}}, \bibinfo {author}
  {\bibfnamefont{F.}~\bibnamefont{Molitor}}, \bibinfo {author}
  {\bibfnamefont{C.}~\bibnamefont{Stampfer}}, \bibinfo {author}
  {\bibfnamefont{J.}~\bibnamefont{Guettinger}}, \bibinfo {author}
  {\bibfnamefont{I.}~\bibnamefont{Shorubalko}}, \bibinfo {author}
  {\bibfnamefont{T.}~\bibnamefont{Ihn}},\ and\ \bibinfo {author}
  {\bibfnamefont{K.}~\bibnamefont{Ensslin}},\ }%
  \bibfield{journal}{%
  \bibinfo {journal} {Appl. Phys. Lett.}\ }%
  \textbf{\bibinfo {volume} {94}},\ \bibinfo {pages} {012107} (\bibinfo {year}
  {2009})%
  \bibAnnoteFile{NoStop}{Setal09b}%
\bibitem{Getal10}%
  \BibitemOpen
  \bibfield{author}{%
  \bibinfo {author} {\bibfnamefont{J.}~\bibnamefont{G\"uttinger}}, \bibinfo
  {author} {\bibfnamefont{T.}~\bibnamefont{Frey}}, \bibinfo {author}
  {\bibfnamefont{C.}~\bibnamefont{Stampfer}}, \bibinfo {author}
  {\bibfnamefont{T.}~\bibnamefont{Ihn}},\ and\ \bibinfo {author}
  {\bibfnamefont{K.}~\bibnamefont{Ensslin}},\ }%
  \bibinfo {howpublished} {arXiv:1002.3771v1} (\bibinfo {year} {2010})%
  \bibAnnoteFile{NoStop}{Getal10}%
\bibitem{Shima93}%
  \BibitemOpen
  \bibfield{author}{%
  \bibinfo {author} {\bibfnamefont{N.}~\bibnamefont{Shima}}\ and\ \bibinfo
  {author} {\bibfnamefont{H.}~\bibnamefont{Aoki}},\ }%
  \bibfield{journal}{%
  \bibinfo {journal} {Phys. Rev. Lett.}\ }%
  \textbf{\bibinfo {volume} {71}},\ \bibinfo {pages} {4389} (\bibinfo {year}
  {1993})%
  \bibAnnoteFile{NoStop}{Shima93}%
\bibitem{Pedersen08}%
  \BibitemOpen
  \bibfield{author}{%
  \bibinfo {author} {\bibfnamefont{T.~G.}\ \bibnamefont{Pedersen}}, \bibinfo
  {author} {\bibfnamefont{C.}~\bibnamefont{Flindt}}, \bibinfo {author}
  {\bibfnamefont{J.}~\bibnamefont{Pedersen}}, \bibinfo {author}
  {\bibfnamefont{N.~A.}\ \bibnamefont{Mortensen}}, \bibinfo {author}
  {\bibfnamefont{A.-P.}\ \bibnamefont{Jauho}},\ and\ \bibinfo {author}
  {\bibfnamefont{K.}~\bibnamefont{Pedersen}},\ }%
  \bibfield{journal}{%
  \bibinfo {journal} {Phys. Rev. Lett.}\ }%
  \textbf{\bibinfo {volume} {100}},\ \bibinfo {pages} {136804} (\bibinfo {year}
  {2008})%
  \bibAnnoteFile{NoStop}{Pedersen08}%
\bibitem{Vanevic09}%
  \BibitemOpen
  \bibfield{author}{%
  \bibinfo {author}
  {\bibfnamefont{M.}~\bibnamefont{Vanevi\ifmmode~\acute{c}\else \'{c}\fi{}}},
  \bibinfo {author} {\bibfnamefont{V.~M.}\
  \bibnamefont{Stojanovi\ifmmode~\acute{c}\else \'{c}\fi{}}},\ and\ \bibinfo
  {author} {\bibfnamefont{M.}~\bibnamefont{Kindermann}},\ }%
  \bibfield{journal}{%
  \bibinfo {journal} {Phys. Rev. B}\ }%
  \textbf{\bibinfo {volume} {80}},\ \bibinfo {pages} {045410} (\bibinfo {year}
  {2009})%
  \bibAnnoteFile{NoStop}{Vanevic09}%
\bibitem{Fuerst09}%
  \BibitemOpen
  \bibfield{author}{%
  \bibinfo {author} {\bibfnamefont{J.~A.}\ \bibnamefont{F\"urst}}, \bibinfo
  {author} {\bibfnamefont{J.~G.}\ \bibnamefont{Pedersen}}, \bibinfo {author}
  {\bibfnamefont{C.}~\bibnamefont{Flindt}}, \bibinfo {author}
  {\bibfnamefont{N.~A.}\ \bibnamefont{Mortensen}}, \bibinfo {author}
  {\bibfnamefont{M.}~\bibnamefont{Brandbyge}}, \bibinfo {author}
  {\bibfnamefont{T.~G.}\ \bibnamefont{Pedersen}},\ and\ \bibinfo {author}
  {\bibfnamefont{A.-P.}\ \bibnamefont{Jauho}},\ }%
  \bibfield{journal}{%
  \bibinfo {journal} {New J. Phys.}\ }%
  \textbf{\bibinfo {volume} {11}},\ \bibinfo {pages} {095020} (\bibinfo {year}
  {2009})%
  \bibAnnoteFile{NoStop}{Fuerst09}%
\bibitem{Shen2008}%
  \BibitemOpen
  \bibfield{author}{%
  \bibinfo {author} {\bibfnamefont{T.}~\bibnamefont{Shen}}, \bibinfo {author}
  {\bibfnamefont{Y.~Q.}\ \bibnamefont{Wu}}, \bibinfo {author}
  {\bibfnamefont{M.~A.}\ \bibnamefont{Capano}}, \bibinfo {author}
  {\bibfnamefont{L.~P.}\ \bibnamefont{Rokhinson}}, \bibinfo {author}
  {\bibfnamefont{L.~W.}\ \bibnamefont{Engel}},\ and\ \bibinfo {author}
  {\bibfnamefont{P.~D.}\ \bibnamefont{Ye}},\ }%
  \bibfield{journal}{%
  \bibinfo {journal} {Appl. Phys. Lett.}\ }%
  \textbf{\bibinfo {volume} {93}},\ \bibinfo {eid} {122102} (\bibinfo {year}
  {2008})%
  \bibAnnoteFile{NoStop}{Shen2008}%
\bibitem{Eroms2009}%
  \BibitemOpen
  \bibfield{author}{%
  \bibinfo {author} {\bibfnamefont{J.}~\bibnamefont{Eroms}}\ and\ \bibinfo
  {author} {\bibfnamefont{D.}~\bibnamefont{Weiss}},\ }%
  \bibfield{journal}{%
  \bibinfo {journal} {New J. Phys.}\ }%
  \textbf{\bibinfo {volume} {11}},\ \bibinfo {pages} {095021} (\bibinfo {year}
  {2009})%
  \bibAnnoteFile{NoStop}{Eroms2009}%
\bibitem{Bai10}%
  \BibitemOpen
  \bibfield{author}{%
  \bibinfo {author} {\bibfnamefont{J.}~\bibnamefont{Bai}}, \bibinfo {author}
  {\bibfnamefont{X.}~\bibnamefont{Zhong}}, \bibinfo {author}
  {\bibfnamefont{S.}~\bibnamefont{Jiang}}, \bibinfo {author}
  {\bibfnamefont{Y.}~\bibnamefont{Huang}},\ and\ \bibinfo {author}
  {\bibfnamefont{X.}~\bibnamefont{Duan}},\ }%
  \bibfield{journal}{%
  \bibinfo {journal} {Nature Nano.}\ }%
  \textbf{\bibinfo {volume} {5}},\ \bibinfo {pages} {190} (\bibinfo {year}
  {2010})%
  \bibAnnoteFile{NoStop}{Bai10}%
\bibitem{Balog10}%
  \BibitemOpen
  \bibfield{author}{%
  \bibinfo {author} {\bibfnamefont{R.}~\bibnamefont{Balog}}, \bibinfo {author}
  {\bibfnamefont{B.}~\bibnamefont{J\o{}rgensen}}, \bibinfo {author}
  {\bibfnamefont{L.}~\bibnamefont{Nilsson}}, \bibinfo {author}
  {\bibfnamefont{M.}~\bibnamefont{Andersen}}, \bibinfo {author}
  {\bibfnamefont{E.}~\bibnamefont{Rienks}}, \bibinfo {author}
  {\bibfnamefont{M.}~\bibnamefont{Bianchi}}, \bibinfo {author}
  {\bibfnamefont{M.}~\bibnamefont{Fanetti}}, \bibinfo {author}
  {\bibfnamefont{E.}~\bibnamefont{L\ae{}gsgaard}}, \bibinfo {author}
  {\bibfnamefont{A.}~\bibnamefont{Baraldi}}, \bibinfo {author}
  {\bibfnamefont{S.}~\bibnamefont{Lizzit}}, \bibinfo {author}
  {\bibfnamefont{Z.}~\bibnamefont{Sljivancanin}}, \bibinfo {author}
  {\bibfnamefont{F.}~\bibnamefont{Besenbacher}}, \bibinfo {author}
  {\bibfnamefont{B.}~\bibnamefont{Hammer}}, \bibinfo {author}
  {\bibfnamefont{T.~G.}\ \bibnamefont{Pedersen}}, \bibinfo {author}
  {\bibfnamefont{P.}~\bibnamefont{Hofmann}},\ and\ \bibinfo {author}
  {\bibfnamefont{L.}~\bibnamefont{Hornek\ae{}r}},\ }%
  \bibfield{journal}{%
  \bibinfo {journal} {Nature Mat.}\ }%
  \textbf{\bibinfo {volume} {9}},\ \bibinfo {pages} {315} (\bibinfo {year}
  {2010})%
  \bibAnnoteFile{NoStop}{Balog10}%
\bibitem{Cancado2004}%
  \BibitemOpen
  \bibfield{author}{%
  \bibinfo {author} {\bibfnamefont{L.~G.}\ \bibnamefont{Cancado}}, \bibinfo
  {author} {\bibfnamefont{M.~A.}\ \bibnamefont{Pimenta}}, \bibinfo {author}
  {\bibfnamefont{B.~R.~A.}\ \bibnamefont{Neves}}, \bibinfo {author}
  {\bibfnamefont{M.~S.~S.}\ \bibnamefont{Dantas}},\ and\ \bibinfo {author}
  {\bibfnamefont{A.}~\bibnamefont{Jorio}},\ }%
  \bibfield{journal}{%
  \Doi{10.1103/PhysRevLett.93.247401}{\bibinfo {journal} {Phys. Rev. Lett.}}\
  }%
  \textbf{\bibinfo {volume} {93}},\ \bibinfo {pages} {247401} (\bibinfo {year}
  {2004})%
  \bibAnnoteFile{NoStop}{Cancado2004}%
\bibitem{Vetal08}%
  \BibitemOpen
  \bibfield{author}{%
  \bibinfo {author} {\bibfnamefont{A.~L.}\ \bibnamefont{{V\'azquez de Parga}}},
  \bibinfo {author} {\bibfnamefont{F.}~\bibnamefont{Calleja}}, \bibinfo
  {author} {\bibfnamefont{B.}~\bibnamefont{Borca}}, \bibinfo {author}
  {\bibfnamefont{M.~C.~G.}\ \bibnamefont{Passeggi}}, \bibinfo {author}
  {\bibfnamefont{J.~J.}\ \bibnamefont{Hinarejos}}, \bibinfo {author}
  {\bibfnamefont{F.}~\bibnamefont{Guinea}},\ and\ \bibinfo {author}
  {\bibfnamefont{R.}~\bibnamefont{Miranda}},\ }%
  \bibfield{journal}{%
  \bibinfo {journal} {Phys. Rev. Lett.}\ }%
  \textbf{\bibinfo {volume} {100}},\ \bibinfo {pages} {056807} (\bibinfo {year}
  {2008})%
  \bibAnnoteFile{NoStop}{Vetal08}%
\bibitem{Jetal09}%
  \BibitemOpen
  \bibfield{author}{%
  \bibinfo {author} {\bibfnamefont{X.}~\bibnamefont{Jia}}, \bibinfo {author}
  {\bibfnamefont{M.}~\bibnamefont{Hofmann}}, \bibinfo {author}
  {\bibfnamefont{V.}~\bibnamefont{Meunier}}, \bibinfo {author}
  {\bibfnamefont{B.~G.}\ \bibnamefont{Sumpter}}, \bibinfo {author}
  {\bibfnamefont{J.}~\bibnamefont{Campos-Delgado}}, \bibinfo {author}
  {\bibfnamefont{J.-M.}\ \bibnamefont{Romo-Herrera}}, \bibinfo {author}
  {\bibfnamefont{H.}~\bibnamefont{Son}}, \bibinfo {author}
  {\bibfnamefont{Y.-P.}\ \bibnamefont{Hsieh}}, \bibinfo {author}
  {\bibfnamefont{A.}~\bibnamefont{Reina}}, \bibinfo {author}
  {\bibfnamefont{J.}~\bibnamefont{Kong}}, \bibinfo {author}
  {\bibfnamefont{M.}~\bibnamefont{Terrones}},\ and\ \bibinfo {author}
  {\bibfnamefont{M.~S.}\ \bibnamefont{Dresselhaus}},\ }%
  \bibfield{journal}{%
  \bibinfo {journal} {Science}\ }%
  \textbf{\bibinfo {volume} {323}},\ \bibinfo {pages} {1701} (\bibinfo {year}
  {2009})%
  \bibAnnoteFile{NoStop}{Jetal09}%
\bibitem{Getal09}%
  \BibitemOpen
  \bibfield{author}{%
  \bibinfo {author} {\bibfnamefont{C.}~\bibnamefont{{\"O. Girit}}}, \bibinfo
  {author} {\bibfnamefont{J.~C.}\ \bibnamefont{Meyer}}, \bibinfo {author}
  {\bibfnamefont{R.}~\bibnamefont{Erni}}, \bibinfo {author}
  {\bibfnamefont{M.~D.}\ \bibnamefont{Rossell}}, \bibinfo {author}
  {\bibfnamefont{C.}~\bibnamefont{Kisielowski}}, \bibinfo {author}
  {\bibfnamefont{L.}~\bibnamefont{Yang}}, \bibinfo {author}
  {\bibfnamefont{C.-H. C.-H.}\ \bibnamefont{Park}}, \bibinfo {author}
  {\bibfnamefont{M.~F.}\ \bibnamefont{Crommie}}, \bibinfo {author}
  {\bibfnamefont{M.~L.}\ \bibnamefont{Cohen}}, \bibinfo {author}
  {\bibfnamefont{S.~G.}\ \bibnamefont{Louie}},\ and\ \bibinfo {author}
  {\bibfnamefont{A.}~\bibnamefont{Zettl}},\ }%
  \bibfield{journal}{%
  \bibinfo {journal} {Science}\ }%
  \textbf{\bibinfo {volume} {323}},\ \bibinfo {pages} {1705} (\bibinfo {year}
  {2009})%
  \bibAnnoteFile{NoStop}{Getal09}%
\bibitem{Liu2009}%
  \BibitemOpen
  \bibfield{author}{%
  \bibinfo {author} {\bibfnamefont{Z.}~\bibnamefont{Liu}}, \bibinfo {author}
  {\bibfnamefont{K.}~\bibnamefont{Suenaga}}, \bibinfo {author}
  {\bibfnamefont{P.~J.~F.}\ \bibnamefont{Harris}},\ and\ \bibinfo {author}
  {\bibfnamefont{S.}~\bibnamefont{Iijima}},\ }%
  \bibfield{journal}{%
  \Doi{10.1103/PhysRevLett.102.015501}{\bibinfo {journal} {Phys. Rev. Lett.}}\
  }%
  \textbf{\bibinfo {volume} {102}},\ \bibinfo {pages} {015501} (\bibinfo {year}
  {2009})%
  \bibAnnoteFile{NoStop}{Liu2009}%
\bibitem{Casiraghi2009}%
  \BibitemOpen
  \bibfield{author}{%
  \bibinfo {author} {\bibfnamefont{C.}~\bibnamefont{Casiraghi}}, \bibinfo
  {author} {\bibfnamefont{A.}~\bibnamefont{Hartschuh}}, \bibinfo {author}
  {\bibfnamefont{H.}~\bibnamefont{Qian}}, \bibinfo {author}
  {\bibfnamefont{S.}~\bibnamefont{Piscanec}}, \bibinfo {author}
  {\bibfnamefont{C.}~\bibnamefont{Georgi}}, \bibinfo {author}
  {\bibfnamefont{A.}~\bibnamefont{Fasoli}}, \bibinfo {author}
  {\bibfnamefont{K.~S.}\ \bibnamefont{Novoselov}}, \bibinfo {author}
  {\bibfnamefont{D.~M.}\ \bibnamefont{Basko}},\ and\ \bibinfo {author}
  {\bibfnamefont{A.~C.}\ \bibnamefont{Ferrari}},\ }%
  \bibfield{journal}{%
  \bibinfo {journal} {Nano Letters}\ }%
  \textbf{\bibinfo {volume} {9}},\ \bibinfo {pages} {1433} (\bibinfo {year}
  {2009})%
  \bibAnnoteFile{NoStop}{Casiraghi2009}%
\bibitem{XL04}%
  \BibitemOpen
  \bibfield{author}{%
  \bibinfo {author} {\bibfnamefont{Y.~J.}\ \bibnamefont{Xu}}\ and\ \bibinfo
  {author} {\bibfnamefont{J.~Q.}\ \bibnamefont{Li}},\ }%
  \bibfield{journal}{%
  \bibinfo {journal} {Chem. Phys. Lett.}\ }%
  \textbf{\bibinfo {volume} {400}},\ \bibinfo {pages} {406} (\bibinfo {year}
  {2004})%
  \bibAnnoteFile{NoStop}{XL04}%
\bibitem{JSD06}%
  \BibitemOpen
  \bibfield{author}{%
  \bibinfo {author} {\bibfnamefont{D.~E.}\ \bibnamefont{Jiang}}, \bibinfo
  {author} {\bibfnamefont{B.~G.}\ \bibnamefont{Sumpter}},\ and\ \bibinfo
  {author} {\bibfnamefont{S.}~\bibnamefont{Dai}},\ }%
  \bibfield{journal}{%
  \bibinfo {journal} {Journ. Phys. Chem. B}\ }%
  \textbf{\bibinfo {volume} {110}},\ \bibinfo {pages} {23628} (\bibinfo {year}
  {2006})%
  \bibAnnoteFile{NoStop}{JSD06}%
\bibitem{CCPF08}%
  \BibitemOpen
  \bibfield{author}{%
  \bibinfo {author} {\bibfnamefont{F.}~\bibnamefont{Cervantes-Sodi}}, \bibinfo
  {author} {\bibfnamefont{G.}~\bibnamefont{{Cs\'anyi}}}, \bibinfo {author}
  {\bibfnamefont{S.}~\bibnamefont{Piscanec}},\ and\ \bibinfo {author}
  {\bibfnamefont{A.~C.}\ \bibnamefont{Ferrari}},\ }%
  \bibfield{journal}{%
  \bibinfo {journal} {Phys. Rev. B}\ }%
  \textbf{\bibinfo {volume} {77}},\ \bibinfo {pages} {165427} (\bibinfo {year}
  {2008})%
  \bibAnnoteFile{NoStop}{CCPF08}%
\bibitem{HLWGD08}%
  \BibitemOpen
  \bibfield{author}{%
  \bibinfo {author} {\bibfnamefont{B.}~\bibnamefont{Huang}}, \bibinfo {author}
  {\bibfnamefont{F.}~\bibnamefont{Liu}}, \bibinfo {author}
  {\bibfnamefont{J.}~\bibnamefont{Wu}}, \bibinfo {author}
  {\bibfnamefont{B.-L.}\ \bibnamefont{Gu}},\ and\ \bibinfo {author}
  {\bibfnamefont{W.}~\bibnamefont{Duan}},\ }%
  \bibfield{journal}{%
  \bibinfo {journal} {Phys. Rev. B}\ }%
  \textbf{\bibinfo {volume} {77}},\ \bibinfo {pages} {153411} (\bibinfo {year}
  {2008})%
  \bibAnnoteFile{NoStop}{HLWGD08}%
\bibitem{footnote_colorplots}%
  \BibitemOpen
  \bibinfo {howpublished} {Close to a boundary, the edge states occupy a single
  sublattice only. In order to avoid the oscillatory pattern on the lattice
  scale that inevitably arises when plotting both sublattices simultaneously,
  for every unit cell we only plot the atom with the largest occupation
  probability.}%
  \bibAnnoteFile{Stop}{footnote_colorplots}%
\bibitem{SESI08}%
  \BibitemOpen
  \bibfield{author}{%
  \bibinfo {author} {\bibfnamefont{S.}~\bibnamefont{Schnez}}, \bibinfo {author}
  {\bibfnamefont{K.}~\bibnamefont{Ensslin}}, \bibinfo {author}
  {\bibfnamefont{M.}~\bibnamefont{Sigrist}},\ and\ \bibinfo {author}
  {\bibfnamefont{T.}~\bibnamefont{Ihn}},\ }%
  \bibfield{journal}{%
  \bibinfo {journal} {Appl. Phys. Lett.}\ }%
  \textbf{\bibinfo {volume} {78}},\ \bibinfo {pages} {195427} (\bibinfo {year}
  {2008})%
  \bibAnnoteFile{NoStop}{SESI08}%
\bibitem{Lieb89}%
  \BibitemOpen
  \bibfield{author}{%
  \bibinfo {author} {\bibfnamefont{E.~H.}\ \bibnamefont{Lieb}},\ }%
  \bibfield{journal}{%
  \bibinfo {journal} {Phys. Rev. Lett.}\ }%
  \textbf{\bibinfo {volume} {62}},\ \bibinfo {pages} {1201} (\bibinfo {year}
  {1989})%
  \bibAnnoteFile{NoStop}{Lieb89}%
\bibitem{Ezawa10}%
  \BibitemOpen
  \bibfield{author}{%
  \bibinfo {author} {\bibfnamefont{M.}~\bibnamefont{Ezawa}},\ }%
  \bibfield{journal}{%
  \bibinfo {journal} {Physica E}\ }%
  \textbf{\bibinfo {volume} {42}},\ \bibinfo {pages} {703 } (\bibinfo {year}
  {2010})%
  \bibAnnoteFile{NoStop}{Ezawa10}%
\bibitem{ABRB08}%
  \BibitemOpen
  \bibfield{author}{%
  \bibinfo {author} {\bibfnamefont{A.~R.}\ \bibnamefont{Akhmerov}}, \bibinfo
  {author} {\bibfnamefont{J.~H.}\ \bibnamefont{Bardarson}}, \bibinfo {author}
  {\bibfnamefont{A.}~\bibnamefont{Rycerz}},\ and\ \bibinfo {author}
  {\bibfnamefont{C.~W.~J.}\ \bibnamefont{Beenakker}},\ }%
  \bibfield{journal}{%
  \bibinfo {journal} {Phys. Rev. B}\ }%
  \textbf{\bibinfo {volume} {77}},\ \bibinfo {pages} {205416} (\bibinfo {year}
  {2008})%
  \bibAnnoteFile{NoStop}{ABRB08}%
\bibitem{PGN06}%
  \BibitemOpen
  \bibfield{author}{%
  \bibinfo {author} {\bibfnamefont{N.~M.~R.}\ \bibnamefont{Peres}}, \bibinfo
  {author} {\bibfnamefont{F.}~\bibnamefont{Guinea}},\ and\ \bibinfo {author}
  {\bibfnamefont{A.~H.}\ \bibnamefont{{Castro Neto}}},\ }%
  \bibfield{journal}{%
  \bibinfo {journal} {Phys. Rev. B}\ }%
  \textbf{\bibinfo {volume} {73}},\ \bibinfo {pages} {125411} (\bibinfo {year}
  {2006})%
  \bibAnnoteFile{NoStop}{PGN06}%
\bibitem{SMS06}%
  \BibitemOpen
  \bibfield{author}{%
  \bibinfo {author} {\bibfnamefont{K.}~\bibnamefont{Sasaki}}, \bibinfo {author}
  {\bibfnamefont{S.}~\bibnamefont{Murakami}},\ and\ \bibinfo {author}
  {\bibfnamefont{R.}~\bibnamefont{Saito}},\ }%
  \bibfield{journal}{%
  \bibinfo {journal} {Appl. Phys. Lett.}\ }%
  \textbf{\bibinfo {volume} {88}},\ \bibinfo {pages} {113110} (\bibinfo {year}
  {2006})%
  \bibAnnoteFile{NoStop}{SMS06}%
\bibitem{Sasaki2009}%
  \BibitemOpen
  \bibfield{author}{%
  \bibinfo {author} {\bibfnamefont{K.-i.}\ \bibnamefont{Sasaki}}, \bibinfo
  {author} {\bibfnamefont{Y.}~\bibnamefont{Shimomura}}, \bibinfo {author}
  {\bibfnamefont{Y.}~\bibnamefont{Takane}},\ and\ \bibinfo {author}
  {\bibfnamefont{K.}~\bibnamefont{Wakabayashi}},\ }%
  \bibfield{journal}{%
  \bibinfo {journal} {Phys. Rev. Lett.}\ }%
  \textbf{\bibinfo {volume} {102}},\ \bibinfo {pages} {146806} (\bibinfo {year}
  {2009})%
  \bibAnnoteFile{NoStop}{Sasaki2009}%
\bibitem{dotshape}%
  \BibitemOpen
  \bibinfo {howpublished} {The dot has the shape of a deformed disk, with a
  radius $R(\theta)$ depending on the angle of direction $\theta$. The
  numerical calculations presented in the text use $R(\theta)=R+0.2 R
  \sin(\theta)+0.05 R \sin(2 \theta)-0.025 R \sin(3 \theta)+0.02 R \sin(4
  \theta)-0.01 R \sin(5 \theta)$}%
  \bibAnnoteFile{NoStop}{dotshape}%
\bibitem{Peierls1933}%
  \BibitemOpen
  \bibfield{author}{%
  \bibinfo {author} {\bibfnamefont{R.}~\bibnamefont{Peierls}},\ }%
  \bibfield{journal}{%
  \Doi{10.1007/BF01342591}{\bibinfo {journal} {Z. f. Phys. A}}\ }%
  \textbf{\bibinfo {volume} {80}},\ \bibinfo {pages} {763} (\bibinfo {month}
  {Nov.}\ \bibinfo {year} {1933})%
  \bibAnnoteFile{NoStop}{Peierls1933}%
\bibitem{Mucciolo2009}%
  \BibitemOpen
  \bibfield{author}{%
  \bibinfo {author} {\bibfnamefont{E.~R.}\ \bibnamefont{Mucciolo}}, \bibinfo
  {author} {\bibfnamefont{A.~H.}\ \bibnamefont{Castro~Neto}},\ and\ \bibinfo
  {author} {\bibfnamefont{C.~H.}\ \bibnamefont{Lewenkopf}},\ }%
  \bibfield{journal}{%
  \Doi{10.1103/PhysRevB.79.075407}{\bibinfo {journal} {Phys. Rev. B}}\ }%
  \textbf{\bibinfo {volume} {79}},\ \bibinfo {eid} {075407} (\bibinfo {year}
  {2009})%
  \bibAnnoteFile{NoStop}{Mucciolo2009}%
\bibitem{Lapack}%
  \BibitemOpen
  \bibfield{author}{%
  \bibinfo {author} {\bibfnamefont{E.}~\bibnamefont{Anderson}}, \bibinfo
  {author} {\bibfnamefont{Z.}~\bibnamefont{Bai}}, \bibinfo {author}
  {\bibfnamefont{C.}~\bibnamefont{Bischof}}, \bibinfo {author}
  {\bibfnamefont{S.}~\bibnamefont{Blackford}}, \bibinfo {author}
  {\bibfnamefont{J.}~\bibnamefont{Demmel}}, \bibinfo {author}
  {\bibfnamefont{J.}~\bibnamefont{Dongarra}}, \bibinfo {author}
  {\bibfnamefont{J.}~\bibnamefont{Du~Croz}}, \bibinfo {author}
  {\bibfnamefont{A.}~\bibnamefont{Greenbaum}}, \bibinfo {author}
  {\bibfnamefont{S.}~\bibnamefont{Hammarling}}, \bibinfo {author}
  {\bibfnamefont{A.}~\bibnamefont{McKenney}},\ and\ \bibinfo {author}
  {\bibfnamefont{D.}~\bibnamefont{Sorensen}},\ }%
  \emph{\bibinfo {title} {{LAPACK} Users' Guide}},\ \bibinfo {edition} {3rd}\
  ed.\ (\bibinfo {publisher} {Society for Industrial and Applied Mathematics},\
  \bibinfo {address} {Philadelphia, PA},\ \bibinfo {year} {1999})%
  \bibAnnoteFile{NoStop}{Lapack}%
\bibitem{Gibbs1976}%
  \BibitemOpen
  \bibfield{author}{%
  \bibinfo {author} {\bibfnamefont{N.~E.}\ \bibnamefont{Gibbs}}, \bibinfo
  {author} {\bibfnamefont{J.}~\bibnamefont{William G.~Poole}},\ and\ \bibinfo
  {author} {\bibfnamefont{P.~K.}\ \bibnamefont{Stockmeyer}},\ }%
  \bibfield{journal}{%
  \Doi{10.1137/0713023}{\bibinfo {journal} {SIAM J. Num. Anal.}}\ }%
  \textbf{\bibinfo {volume} {13}},\ \bibinfo {pages} {236} (\bibinfo {year}
  {1976})%
  \bibAnnoteFile{NoStop}{Gibbs1976}%
\bibitem{ARPACK}%
  \BibitemOpen
  \bibfield{author}{%
  \bibinfo {author} {\bibfnamefont{R.~B.}\ \bibnamefont{Lehoucq}}, \bibinfo
  {author} {\bibfnamefont{D.~C.}\ \bibnamefont{Sorensen}},\ and\ \bibinfo
  {author} {\bibfnamefont{C.}~\bibnamefont{Yang}},\ }%
  \emph{\bibinfo {title} {ARPACK Users' Guide: Solution of Large-Scale
  Eigenvalue Problems with Implicitly Restarted Arnoldi Methods}}\ (\bibinfo
  {publisher} {Society for Industrial and Applied Mathematics},\ \bibinfo
  {address} {Philadelphia, PA},\ \bibinfo {year} {1998})%
  \bibAnnoteFile{NoStop}{ARPACK}%
\bibitem{MUMPS}%
  \BibitemOpen
  \bibinfo {howpublished} {For the solution of the sparse linear system arising
  in the shift-and-invert problem, we apply the MUMPS package: P. R. Amestoy,
  I. S. Duff, J. Koster, and J.-Y. L'Excellent, SIAM J. Matrix Anal. Appl. {\bf
  23}, 15 (2001).}%
  \bibAnnoteFile{Stop}{MUMPS}%
\bibitem{Sasaki2007b}%
  \BibitemOpen
  \bibfield{author}{%
  \bibinfo {author} {\bibfnamefont{K.}~\bibnamefont{Sasaki}}, \bibinfo {author}
  {\bibfnamefont{K.}~\bibnamefont{Sato}}, \bibinfo {author}
  {\bibfnamefont{R.}~\bibnamefont{Saito}}, \bibinfo {author}
  {\bibfnamefont{J.}~\bibnamefont{Jiang}}, \bibinfo {author}
  {\bibfnamefont{S.}~\bibnamefont{Onari}},\ and\ \bibinfo {author}
  {\bibfnamefont{Y.}~\bibnamefont{Tanaka}},\ }%
  \bibfield{journal}{%
  \bibinfo {journal} {Phys. Rev. B}\ }%
  \textbf{\bibinfo {volume} {75}},\ \bibinfo {pages} {235430} (\bibinfo {year}
  {2007})%
  \bibAnnoteFile{NoStop}{Sasaki2007b}%
\bibitem{Bahamon09}%
  \BibitemOpen
  \bibfield{author}{%
  \bibinfo {author} {\bibfnamefont{D.~A.}\ \bibnamefont{Bahamon}}, \bibinfo
  {author} {\bibfnamefont{A.~L.~C.}\ \bibnamefont{Pereira}},\ and\ \bibinfo
  {author} {\bibfnamefont{P.~A.}\ \bibnamefont{Schulz}},\ }%
  \bibfield{journal}{%
  \bibinfo {journal} {Phys. Rev. B}\ }%
  \textbf{\bibinfo {volume} {79}},\ \bibinfo {pages} {125414} (\bibinfo {year}
  {2009})%
  \bibAnnoteFile{NoStop}{Bahamon09}%
\bibitem{Kim10}%
  \BibitemOpen
  \bibfield{author}{%
  \bibinfo {author} {\bibfnamefont{S.~C.}\ \bibnamefont{Kim}}, \bibinfo
  {author} {\bibfnamefont{P.~S.}\ \bibnamefont{Park}},\ and\ \bibinfo {author}
  {\bibfnamefont{S.-R.~E.}\ \bibnamefont{Yang}},\ }%
  \bibfield{journal}{%
  \bibinfo {journal} {Phys. Rev. B}\ }%
  \textbf{\bibinfo {volume} {81}},\ \bibinfo {pages} {085432} (\bibinfo {year}
  {2010})%
  \bibAnnoteFile{NoStop}{Kim10}%
\bibitem{Bell1970}%
  \BibitemOpen
  \bibfield{author}{%
  \bibinfo {author} {\bibfnamefont{R.~J.}\ \bibnamefont{Bell}}\ and\ \bibinfo
  {author} {\bibfnamefont{P.}~\bibnamefont{Dean}},\ }%
  \bibfield{journal}{%
  \Doi{10.1039/DF9705000055}{\bibinfo {journal} {Discuss. Faraday Soc.}}\ }%
  \textbf{\bibinfo {volume} {50}},\ \bibinfo {pages} {55} (\bibinfo {year}
  {1970})%
  \bibAnnoteFile{NoStop}{Bell1970}%
\bibitem{Bell1972}%
  \BibitemOpen
  \bibfield{author}{%
  \bibinfo {author} {\bibfnamefont{R.~J.}\ \bibnamefont{Bell}},\ }%
  \bibfield{journal}{%
  \Doi{10.1088/0034-4885/35/3/306}{\bibinfo {journal} {Rep. Prog. Phys.}}\ }%
  \textbf{\bibinfo {volume} {35}},\ \bibinfo {pages} {1315} (\bibinfo {year}
  {1972})%
  \bibAnnoteFile{NoStop}{Bell1972}%
\bibitem{Libisch2010}%
  \BibitemOpen
  \bibfield{author}{%
  \bibinfo {author} {\bibfnamefont{F.}~\bibnamefont{Libisch}}, \bibinfo
  {author} {\bibfnamefont{S.}~\bibnamefont{Rotter}}, \bibinfo {author}
  {\bibfnamefont{J.}~\bibnamefont{G\"uttinger}}, \bibinfo {author}
  {\bibfnamefont{C.}~\bibnamefont{Stampfer}},\ and\ \bibinfo {author}
  {\bibfnamefont{J.}~\bibnamefont{Burgd\"orfer}},\ }%
  \bibfield{journal}{%
  \Doi{10.1103/PhysRevB.81.245411}{\bibinfo {journal} {Phys. Rev. B}}\ }%
  \textbf{\bibinfo {volume} {81}},\ \bibinfo {pages} {245411} (\bibinfo {year}
  {2010})%
  \bibAnnoteFile{NoStop}{Libisch2010}%
\bibitem{Wetal09}%
  \BibitemOpen
  \bibfield{author}{%
  \bibinfo {author} {\bibfnamefont{J.}~\bibnamefont{Wurm}}, \bibinfo {author}
  {\bibfnamefont{A.}~\bibnamefont{Rycerz}}, \bibinfo {author}
  {\bibfnamefont{I.}~\bibnamefont{Adagideli}}, \bibinfo {author}
  {\bibfnamefont{M.}~\bibnamefont{Wimmer}}, \bibinfo {author}
  {\bibfnamefont{K.}~\bibnamefont{Richter}},\ and\ \bibinfo {author}
  {\bibfnamefont{H.~U.}\ \bibnamefont{Baranger}},\ }%
  \bibfield{journal}{%
  \bibinfo {journal} {Phys. Rev. Lett.}\ }%
  \textbf{\bibinfo {volume} {102}},\ \bibinfo {pages} {056806} (\bibinfo {year}
  {2009})%
  \bibAnnoteFile{NoStop}{Wetal09}%
\bibitem{Libisch09}%
  \BibitemOpen
  \bibfield{author}{%
  \bibinfo {author} {\bibfnamefont{F.}~\bibnamefont{Libisch}}, \bibinfo
  {author} {\bibfnamefont{C.}~\bibnamefont{Stampfer}},\ and\ \bibinfo {author}
  {\bibfnamefont{J.}~\bibnamefont{Burgd\"orfer}},\ }%
  \bibfield{journal}{%
  \bibinfo {journal} {Phys. Rev. B}\ }%
  \textbf{\bibinfo {volume} {79}},\ \bibinfo {pages} {115423} (\bibinfo {year}
  {2009})%
  \bibAnnoteFile{NoStop}{Libisch09}%
\bibitem{Mehta}%
  \BibitemOpen
  \bibfield{author}{%
  \bibinfo {author} {\bibfnamefont{M.~L.}\ \bibnamefont{Mehta}},\ }%
  \emph{\bibinfo {title} {Random Matrices}}\ (\bibinfo {publisher}
  {Elsevier/Academic Press},\ \bibinfo {address} {Amsterdam},\ \bibinfo {year}
  {2004})%
  \bibAnnoteFile{NoStop}{Mehta}%
\bibitem{footnote_idedge}%
  \BibitemOpen
  \bibinfo {howpublished} {The result does not change qualitatively if the
  threshhold participation ratio is changed. In particular, the classification
  into the different types of level statistics does not depend on this
  threshold.}%
  \bibAnnoteFile{Stop}{footnote_idedge}%
\bibitem{BF06}%
  \BibitemOpen
  \bibfield{author}{%
  \bibinfo {author} {\bibfnamefont{L.}~\bibnamefont{Brey}}\ and\ \bibinfo
  {author} {\bibfnamefont{H.~A.}\ \bibnamefont{Fertig}},\ }%
  \bibfield{journal}{%
  \bibinfo {journal} {Phys. Rev. B}\ }%
  \textbf{\bibinfo {volume} {73}},\ \bibinfo {pages} {235411} (\bibinfo {year}
  {2006})%
  \bibAnnoteFile{NoStop}{BF06}%
\bibitem{footnote_oldpaper}%
  \BibitemOpen
  \bibinfo {howpublished} {In Ref.~\onlinecite{Wetal09} the chiral symmetry was
  broken by a mass term.}%
  \bibAnnoteFile{Stop}{footnote_oldpaper}%
\bibitem{H01}%
  \BibitemOpen
  \bibfield{author}{%
  \bibinfo {author} {\bibfnamefont{K.}~\bibnamefont{Harigaya}},\ }%
  \bibfield{journal}{%
  \bibinfo {journal} {Chem. Phys. Lett.}\ }%
  \textbf{\bibinfo {volume} {340}},\ \bibinfo {pages} {123} (\bibinfo {year}
  {2001})%
  \bibAnnoteFile{NoStop}{H01}%
\bibitem{HE02}%
  \BibitemOpen
  \bibfield{author}{%
  \bibinfo {author} {\bibfnamefont{K.}~\bibnamefont{Harigaya}}\ and\ \bibinfo
  {author} {\bibfnamefont{T.}~\bibnamefont{Enoki}},\ }%
  \bibfield{journal}{%
  \bibinfo {journal} {Chem. Phys. Lett.}\ }%
  \textbf{\bibinfo {volume} {351}},\ \bibinfo {pages} {128} (\bibinfo {year}
  {2002})%
  \bibAnnoteFile{NoStop}{HE02}%
\bibitem{BFS07}%
  \BibitemOpen
  \bibfield{author}{%
  \bibinfo {author} {\bibfnamefont{L.}~\bibnamefont{Brey}}, \bibinfo {author}
  {\bibfnamefont{H.~A.}\ \bibnamefont{Fertig}},\ and\ \bibinfo {author}
  {\bibfnamefont{S.}~\bibnamefont{Das~Sarma}},\ }%
  \bibfield{journal}{%
  \bibinfo {journal} {Phys. Rev. Lett.}\ }%
  \textbf{\bibinfo {volume} {99}},\ \bibinfo {pages} {116802} (\bibinfo {year}
  {2007})%
  \bibAnnoteFile{NoStop}{BFS07}%
\bibitem{Martin2008}%
  \BibitemOpen
  \bibfield{author}{%
  \bibinfo {author} {\bibfnamefont{J.}~\bibnamefont{Martin}}, \bibinfo {author}
  {\bibfnamefont{N.}~\bibnamefont{Akerman}}, \bibinfo {author}
  {\bibfnamefont{G.}~\bibnamefont{Ulbricht}}, \bibinfo {author}
  {\bibfnamefont{T.}~\bibnamefont{Lohmann}}, \bibinfo {author}
  {\bibfnamefont{J.~H.}\ \bibnamefont{Smet}}, \bibinfo {author}
  {\bibfnamefont{K.}~\bibnamefont{von Klitzing}},\ and\ \bibinfo {author}
  {\bibfnamefont{A.}~\bibnamefont{Yacoby}},\ }%
  \bibfield{journal}{%
  \bibinfo {journal} {Nature Phys.}\ }%
  \textbf{\bibinfo {volume} {4}},\ \bibinfo {pages} {144} (\bibinfo {year}
  {2008})%
  \bibAnnoteFile{NoStop}{Martin2008}%
\bibitem{Lee08}%
  \BibitemOpen
  \bibfield{author}{%
  \bibinfo {author} {\bibfnamefont{E.~H.}\ \bibnamefont{Lee}}, \bibinfo
  {author} {\bibfnamefont{K.}~\bibnamefont{Balasubramanian}}, \bibinfo {author}
  {\bibfnamefont{R.~T.}\ \bibnamefont{Weitz}}, \bibinfo {author}
  {\bibfnamefont{M.}~\bibnamefont{Burghard}},\ and\ \bibinfo {author}
  {\bibfnamefont{K.}~\bibnamefont{Kern}},\ }%
  \bibfield{journal}{%
  \bibinfo {journal} {Nature Nano.}\ }%
  \textbf{\bibinfo {volume} {3}},\ \bibinfo {pages} {486} (\bibinfo {year}
  {2008})%
  \bibAnnoteFile{NoStop}{Lee08}%
\bibitem{Hey10}%
  \BibitemOpen
  \bibfield{author}{%
  \bibinfo {author} {\bibfnamefont{S.}~\bibnamefont{Heydrich}}, \bibinfo
  {author} {\bibfnamefont{M.}~\bibnamefont{Hirmer}}, \bibinfo {author}
  {\bibfnamefont{C.}~\bibnamefont{Preis}}, \bibinfo {author}
  {\bibfnamefont{T.}~\bibnamefont{Korn}}, \bibinfo {author}
  {\bibfnamefont{J.}~\bibnamefont{Eroms}}, \bibinfo {author}
  {\bibfnamefont{D.}~\bibnamefont{Weiss}},\ and\ \bibinfo {author}
  {\bibfnamefont{C.}~\bibnamefont{Sch\"uller}},\ }%
  \bibinfo {howpublished} {arXiv:1006.2067v1} (\bibinfo {year} {2010})%
  \bibAnnoteFile{NoStop}{Hey10}%
\bibitem{RK08}%
  \BibitemOpen
  \bibfield{author}{%
  \bibinfo {author} {\bibfnamefont{H.~D.}\ \bibnamefont{Raedt}}\ and\ \bibinfo
  {author} {\bibfnamefont{M.~I.}\ \bibnamefont{Katsnelson}},\ }%
  \bibfield{journal}{%
  \bibinfo {journal} {JETP Letters}\ }%
  \textbf{\bibinfo {volume} {88}},\ \bibinfo {pages} {607} (\bibinfo {year}
  {2008})%
  \bibAnnoteFile{NoStop}{RK08}%
\end{thebibliography}
\end{document}